\documentclass[aps,prd,superscriptaddress,nofootinbib,floats,preprint]{revtex4-1}
\usepackage{bm}
\usepackage{soul}
\usepackage{ulem}
\usepackage{indentfirst}
\usepackage{amsmath}
\usepackage{graphicx}
\usepackage{float}
\usepackage{amssymb}
\usepackage{subfigure}
\usepackage{amssymb}
\usepackage{psfrag}
\usepackage{hyperref}
\usepackage{epstopdf}
\usepackage{tensor}
\usepackage{comment}
\hypersetup{
  colorlinks=true,
  linkcolor=red,
  citecolor=blue,
}

\begin{document}

\title{On the waveform of the scalar induced gravitational waves}
\author{Fengge Zhang}
\email{fenggezhang@hust.edu.cn}
\affiliation{School of Physics, Huazhong University of Science and Technology,
Wuhan, Hubei 430074, China}
\author{Arshad Ali}
\email{aa$_$math@hust.edu.cn}
\affiliation{School of Physics, Huazhong University of Science and Technology,
Wuhan, Hubei 430074, China}
\author{Yungui Gong}
\email{Corresponding author. yggong@hust.edu.cn}
\affiliation{School of Physics, Huazhong University of Science and Technology,
Wuhan, Hubei 430074, China}
\author{Jiong Lin}
\email{jionglin@hust.edu.cn}
\affiliation{School of Physics, Huazhong University of Science and Technology,
Wuhan, Hubei 430074, China}
\author{Yizhou Lu}
\email{louischou@hust.edu.cn}
\affiliation{School of Physics, Huazhong University of Science and Technology,
Wuhan, Hubei 430074, China}

\begin{abstract}
The scalar induced gravitational waves (SIGWs) is a useful tool to probe the physics in the early universe.
To study inflationary models with this tool, we need to
know how the waveform of SIGWs is related to the shape of the scalar power spectrum.
We propose two parameterizations to approximate the scalar power
spectrum with either a sharp or a broad spike at small scales,
and then use these two parameterizations to study the relation between the shapes of $\Omega_{GW}$ and the scalar power spectrum.
We find that the waveform of SIGWs has a similar shape to the power spectrum.
Away from the peak of the spike, the frequency relation $\Omega_{GW}(k)\sim \mathcal{P}_\zeta^2(k)$ holds
independent of the functional form of the scalar power spectrum. We also give a physical explanation for this general relationship.
The general relation is useful for determining the scalar power spectrum and probing inflationary physics with the waveform of SIGWs.

\end{abstract}

\maketitle

\section{Introduction}
After the detection of gravitational waves (GWs) by the Laser Interferometer Gravitational-Wave Observatory (LIGO) scientific collaboration and Virgo collaboration \cite{Abbott:2016nmj,Abbott:2016blz,Abbott:2017gyy,TheLIGOScientific:2017qsa,Abbott:2017oio,Abbott:2017vtc,LIGOScientific:2018mvr,
Abbott:2020khf,Abbott:2020uma,LIGOScientific:2020stg}, a new window to probe the property of gravity in the strong field and nonlinear regions was opened.
Meanwhile, the idea that primordial black holes (PBHs) \cite{Carr:1974nx,Hawking:1971ei} might consist of a considerable fraction of dark matter (DM) has attracted more and more interests \cite{Ivanov:1994pa,Frampton:2010sw,Belotsky:2014kca,Khlopov:2004sc,
Clesse:2015wea,Carr:2016drx,Pi:2017gih,Inomata:2017okj,Garcia-Bellido:2017fdg,Kovetz:2017rvv} as the confirmed black hole binary for the first GW event GW150916 might be PBHs \cite{Bird:2016dcv,Sasaki:2016jop}. PBH DM may even be used to explain the Planet 9 \cite{Scholtz:2019csj}. To produce significant PBHs as DM, the scalar power spectrum generated during inflation in the early universe should be amplified from ${\mathcal{O}\left(10^{-9}\right)}$ at large scales, at least, seven orders of magnitudes to ${\mathcal{O}\left(10^{-2}\right)}$ at small scales \cite{Gong:2017qlj,Motohashi:2017kbs,Lu:2019sti,Sato-Polito:2019hws}.
With the large power spectrum, associated with the production of PBHs, the scalar induced GWs (SIGWs) at the second order \cite{Bugaev:2009zh,Bugaev:2010bb,Alabidi:2012ex,Inomata:2016rbd,Drees:2019xpp,Inomata:2019ivs,Inomata:2019zqy,DeLuca:2019llr,Kuroyanagi:2018csn,Nakama:2016gzw,Saito:2009jt,Saito:2008jc,Mollerach:2003nq} are also produced which may be detected by the space-based GW observatory like Laser Interferometer Space Antenna (LISA) \cite{Audley:2017drz,Danzmann:1997hm}, TianQin \cite{Luo:2015ght} and Taiji \cite{Hu:2017mde}, the Pulsar Timing Array (PTA) \cite{Hobbs:2013aka,McLaughlin:2013ira,Hobbs:2009yy,Kramer:2013kea} and the Square Kilometer Array (SKA) \cite{Moore:2014lga} in the future.
Therefore we can use SIGWs to probe the physics in the early universe at small scales.

In usual slow-roll inflationary models with a single
canonical scalar field, it is impossible to enhance the power spectrum large enough to produce a significant abundance of PBH DM \cite{Motohashi:2017kbs,Bezrukov:2017dyv,Passaglia:2018ixg}.
In order to amplify the power spectrum by seven orders of magnitude at small scales,
the slow-roll conditions must be violated.
It becomes a challenge to fine-tune the model parameters to enhance the
amplitude of the primordial curvature perturbation to the order of
$\mathcal{O}(0.01)$ while keeping the total number of e-folds
to be $N\simeq 50-60$ and satisfying the CMB constraints \cite{Sasaki:2018dmp,Passaglia:2018ixg}.
Recently there are some progresses toward building successful models
\cite{Garcia-Bellido:2017mdw,Ezquiaga:2017fvi,Dimopoulos:2017ged,Espinosa:2017sgp,
Espinosa:2018eve,Ballesteros:2017fsr,Germani:2017bcs,Gong:2017qlj,Motohashi:2017kbs,Cheng:2018qof,
Gao:2018pvq,Cicoli:2018asa,Passaglia:2018ixg,
Bhaumik:2019tvl,Drees:2019xpp,Xu:2019bdp, Fu:2019ttf,Passaglia:2019ueo,Lin:2020goi,Palma:2020ejf,Braglia:2020eai,Yi:2020kmq}.

Due to the difficulty in constructing  inflationary models,
some phenomenologically primordial scalar power spectrums were usually used to
study the production of PBHs and the waveform of SIGWs \cite{Ananda:2006af,Saito:2008jc,Namba:2015gja,Nakama:2015nea,Nakama:2016enz,Bartolo:2018evs,Byrnes:2018txb,Orlofsky:2016vbd, Inomata:2016rbd,Garcia-Bellido:2017aan,Kohri:2018awv,Bartolo:2018rku, Cai:2018dig,Lu:2019sti,Unal:2018yaa,Inomata:2018epa,Germani:2018jgr,DeLuca:2019llr,Pi:2020otn}.
The simplest parameterization for a peak is a ${\delta}$ function \cite{Ananda:2006af,Saito:2008jc,Namba:2015gja,Nakama:2015nea,Nakama:2016enz,Inomata:2016rbd,Bartolo:2018evs,Byrnes:2018txb}. For
this monochromatic power spectrum, SIGWs can be calculated analytically
and the corresponding fractional energy density ${\Omega_{GW}}$ has two peaks.
Other widely used parameterizations of the primordial scalar power spectrum include the so-called lognormal peak
that reduce to the ${\delta}$ function when the width of the peak ${\sigma}$ tends to zero \cite{Inomata:2016rbd,Lu:2019sti} and Gaussian functions \cite{Namba:2015gja,Garcia-Bellido:2017aan,Cai:2018dig,Germani:2018jgr,Lu:2019sti}.
To better understand the waveform of SIGWs and the relation
between the shapes of SIGWs and the scalar power spectrum,
we propose two models to parameterize the scalar power spectrum
with either a sharp or a broad spike.
The understanding of the relation helps identify the mechanism
of the production of SIGWs.

This paper is organized as follows. In section II,
we review the basic formulas to calculate the SIGWs.
In section III, we propose two parameterizations to approximate
the scalar power spectrum with either a sharp or a broad spike at small scales, respectively.
Then we compute SIGWs with these power spectra and study the general relationship between the waveform of SIGWs and the power spectrum.
Our conclusion is drawn in section IV.

\section{The production of SIGWs}

Under Newtonian gauge, the perturbed metric is
\begin{equation}
ds^2=a^2(\eta)\left[-(1+2\Phi)d\eta^2+\left\{(1-2\Phi)\delta_{ij}+\frac{1}{2}h_{ij}\right\}dx^i x^j\right],
\end{equation}
where ${\Phi}$ is the Bardeen potential, and the second order tensor perturbation ${h_{ij}}$ is transverse and traceless, ${\partial_{i}h_{ij}=h_{ii}=0}$. In Fourier space, the tensor perturbation is
\begin{equation}
h_{ij}\left(\bm{x},\eta\right)=\int\frac{d^3\bm{k}}{\left(2\pi\right)^{3/2}}\text{e}^{i\bm{k}\cdot\bm{x}}\left[h^{+}_{\bm{k}}\left(\eta\right)e^{+}_{ij}\left(\bm{k}\right)
+h^{\times}_{\bm{k}}\left(\eta\right)e^{\times}_{ij}\left(\bm{k}\right)\right],
\end{equation}
where the polarization tensors ${e^{+}_{ij}\left(\bm{k}\right)}$ and ${e^{\times}_{ij}\left(\bm{k}\right)}$ are
\begin{equation}
\begin{split}
 &e^{+}_{ij}\left(\bm{k}\right)=\frac{1}{\sqrt{2}}\left[e_i\left(\bm{k}\right)e_j\left(\bm{k}\right)
 -\overline{e}_i\left(\bm{k}\right)\overline{e}_j\left(\bm{k}\right)\right],\\&
 e^{\times}_{ij}\left(\bm{k}\right)=\frac{1}{\sqrt{2}}\left[e_i\left(\bm{k}\right)
 \overline{e}_j\left(\bm{k}\right)+\overline{e}_i\left(\bm{k}\right)e_j\left(\bm{k}\right)\right],
 \end{split}
\end{equation}
${e_{i}\left(\bm{k}\right)}$ and ${\overline{e}_{i}\left(\bm{k}\right)}$ are two orthonormal basis vectors which are orthogonal to the wave vector ${\bm{k}}$.

Neglecting the anisotropic stress, the equation of motion for the tensor perturbation with either polarization in Fourier space is
\begin{equation}\label{E4}
    h^{''}_{\bm{k}}\left(\eta\right)+2\mathcal{H}h^{'}_{\bm{k}}\left(\eta\right)+k^2h^{}_{\bm{k}}\left(\eta\right)=4S_{\bm{k}}\left(\eta\right),
\end{equation}
where ${\mathcal{H}=a'/a}$, the prime denotes derivative with respect to the conformal time $\tau$, the source ${S_{\bm{k}}\left(\eta\right)=-e^{ij}\left(\bm{k}\right)S_{ij}\left(\eta,\bm{k}\right)}$ is
\begin{equation}\label{ES}
S_{\bm{k}}\left(\eta\right)=\int\frac{d^3\tilde{\bm{k}}}{\left(2\pi\right)^{3/2}}
e^{ij}\left(\bm{k}\right)\tilde{k}_i\tilde{k}_j
f(\bm{k},\tilde{\bm{k}},\eta)\phi_{\tilde{\bm{k}}}\phi_{\bm{k}-\tilde{\bm{k}}},
\end{equation}
\begin{equation}
\begin{split}
f(\bm{k},\tilde{\bm{k}},\eta)=&
2T_{\phi}(\tilde{k}\eta)T_{\phi}(|\bm{k}-\tilde{\bm{k}}|\eta)\\
&+\frac{4}{3\left(1+w\right)\mathcal{H}^2}
\left(T'_{\phi}(\tilde{k}\eta)+\mathcal{H}
T_{\phi}(\tilde{k}\eta)\right)\left(T'_{\phi}(|\bm{k}-\tilde{\bm{k}}|\eta)
+\mathcal{H} T_{\phi}(|\bm{k}-\tilde{\bm{k}}|\eta)\right),
\end{split}
\end{equation}
$e_{ij}^+\bm({k})\tilde{k}^i\tilde{k}^j=\tilde{k}^2[1-\mu^2]\cos(2\varphi)/\sqrt{2}$, $e_{ij}^\times\bm({k})\tilde{k}^i\tilde{k}^j=\tilde{k}^2[1-\mu^2]\sin(2\varphi)/\sqrt{2}$,
$\mu=\bm{k}\cdot\tilde{\bm{k}}/(k\tilde{k})$, $\varphi$ is the azimuthal angle and the transfer function $T_\phi$ for the Bardeen potential $\Phi$ is defined with its primordial value $\phi_{\bm{k}}$ as
\begin{equation}
    \Phi_{\bm{k}}\left(\eta\right)=\phi_{\bm{k}}T_{\phi}\left(k\eta\right).
\end{equation}
The primordial value ${\phi_{\bm{k}}}$ is related with the primordial curvature power spectrum as
\begin{equation}\label{Eps}
    \langle\phi_{\bm{k}}\phi_{\bm{p}}\rangle=\delta^3\left(\bm{k}+\bm{p}\right)\frac{2\pi^2}{k^3}\left(\frac{3+3w}{5+3w}\right)^2
    \mathcal{P}_{\zeta}\left(k\right).
\end{equation}
The solution to Eq. \eqref{E4} is
\begin{equation}\label{E7}
h_{\bm{k}}\left(\eta\right)=\frac{4}{a\left(\eta\right)}\int^{\eta}d\overline{\eta}G_{k}\left(\eta,\overline{\eta}\right)
a\left(\overline{\eta}\right)
S_{\bm{k}}\left(\overline{\eta}\right),
\end{equation}
where the Green's function satisfies the equation
\begin{equation}
    G^{''}_{\bm{k}}+\left(k^2-\frac{a^{''}\left(\eta\right)}{a\left(\eta\right)}\right)G_{\bm{k}}\left(\eta,\overline{\eta}\right)=
    \delta\left(\eta-\overline{\eta}\right).
\end{equation}

With the solution for $h_{\bm{k}}$, we can compute the power spectrum of SIGWs
\begin{equation}\label{Ep}
\langle h_{\bm{k}}\left(\eta\right)h_{\bm{p}}\left(\eta\right)\rangle
=\delta^3\left(\bm{k}+\bm{p}\right) \frac{2\pi^2}{k^3}\mathcal{P}_h\left(k,\eta\right).
\end{equation}

Combining Eqs. \eqref{ES}, \eqref{E7} and \eqref{Ep}, we get
\begin{equation}\label{Eph1}
   \mathcal{P}_h\left(k,\eta\right)=4\int^{\infty}_{0} d\tilde{k}\int^{1}_{-1}d\mu\left(1-\mu^2\right)^2\mathcal{I}^2_{RD}(\bm{k},\bm{k}-\tilde{\bm{k}},\eta)
   \frac{k^3\tilde{k}^3}{|\bm{k}-\tilde{\bm{k}}|^3}\mathcal{P}_{\zeta}(\tilde{k})\mathcal{P}_{\zeta}
   (|\bm{k}-\tilde{\bm{k}}|),
\end{equation}
where
\begin{equation}
\mathcal{I}_{RD}(\bm{k},\bm{k}-\tilde{\bm{k}},\eta)=\left(\frac{3+3w}{5+3w}\right)^2
\frac{4}{a(\eta)}\int^{\eta}_0 d\overline{\eta}G_{k}\left(\eta,\overline{\eta}\right)
a\left(\overline{\eta}\right)f(\bm{k},\tilde{\bm{k}},\overline{\eta}),
\end{equation}
and SIGWs are assumed to be produced before the horizon reentry.
During radiation domination, ${w=1/3}$, the Green's function is
\begin{equation}
    G_{\bm{k}}\left(\eta,\overline{\eta}\right)=\frac{\sin\left[k\left(\eta-\overline{\eta}\right)\right]}{k},
\end{equation}
and tne transfer function is
\begin{equation}
    T_{\phi}\left(x\right)=\frac{9}{x^2}\left(\frac{\sin\left(x/\sqrt{3}\right)}{x/\sqrt{3}}-\cos\left(x/\sqrt{3}\right)\right),
\end{equation}
where ${x=k\eta}$. After some calculations,
we get the power spectrum of the SIGWs \cite{Kohri:2018awv,Lu:2019sti,Baumann:2007zm,Inomata:2016rbd,Espinosa:2018eve,Ananda:2006af}
\begin{equation}\label{Eph2}
\mathcal{P}_{h}\left(k,\eta\right)=4\int^{\infty}_{0}dv\int^{1+v}_{\left|1-v\right|}du \left(\frac{4v^2-\left(1+v^2-u^2\right)^2}{4vu}\right)^2
I^{2}_{RD}\left(u,v,x\right)\mathcal{P}_\zeta\left(vk\right)\mathcal{P}_\zeta\left(uk\right),
\end{equation}
where ${u=|\bm{k}-\tilde{\bm{k}}|/k, v=\tilde{|\bm{k}|}/k}, I_{RD}(u,v,x)=k^2 \mathcal{I}_{RD}(\bm{k},\bm{k}-\tilde{\bm{k}},\eta)$.

 At late time, ${k\eta \gg 1,x\rightarrow \infty}$, the time average of ${I^{2}_{RD}\left(u,v,x\rightarrow\infty \right)}$ is \cite{Kohri:2018awv}

\begin{equation}\label{IRD}
\begin{split}
\overline{I_{\mathrm{RD}}^2(v,u,x\rightarrow \infty)} =& \frac{1}{2x^2} \left( \frac{3(u^2+v^2-3)}{4 u^3 v^3 } \right)^2 \left\{ \left( -4uv+(u^2+v^2-3) \log \left| \frac{3-(u+v)^2}{3-(u-v)^2} \right| \right)^2  \right. \\
&  \left.\vphantom{\log \left| \frac{3-(u+v)^2}{3-(u-v)^2} \right|^2} \qquad \qquad \qquad \qquad \qquad   + \pi^2 (u^2+v^2-3)^2 \Theta \left( v+u-\sqrt{3}\right)\right\}.
\end{split}
\end{equation}

The density parameter of SIGWs per logarithmic interval of ${k}$ is
 \begin{equation}\label{EGW0}
\Omega_{GW}\left(\eta,k\right)=\frac{1}{24}\left(\frac{k}{\mathcal{H\left(\eta\right)}}\right)^2\overline{\mathcal{P}_h\left(k,\eta\right)}.
\end{equation}

Because the energy density of SIGWs decays like radiation, so we can estimate the energy density of SIGWs today in terms of the present  energy density of radiation  ${\Omega_{r,0}}$ as \cite{Espinosa:2018eve}
\begin{equation}\label{EGW}
\Omega_{GW}\left(\eta_0,k\right)=\Omega_{GW}\left(\eta,k\right)\frac{\Omega_{r,0}}{\Omega_{r}\left(\eta\right)},
\end{equation}
where ${\eta}$ can be chosen at a generic time towarding the end of the radiation domination era.

\section{The parameterizations of the power spectrum}
The waveform of the stochastic GW background is usually approximated as a power law in frequency $\Omega_{GW}\sim \Omega_n(f/f_p)^n$,
and the spectral index $n$ is the characterization of a class of theoretical models \cite{Kuroyanagi:2018csn}.
For primordial GWs produced by slow-roll inflation, $n=-2$ in the frequency
range $3\times 10^{-18}-1\times 10^{-16}$ Hz and $n\approx 0$ when the frequency is above $10^{-16}$ Hz.
For stochastic GWs produced by coalescing binary system, $n=2/3$ before the peak frequency $f_p$.
For GWs generated from domain walls, $n=3$ before the peak frequency $f_p$
and $n=-1$ after the peak frequency \cite{Hiramatsu:2013qaa}.
Therefore, it is important to study the correspondence between the waveform of SIGWs and the scalar power spectrum.
We use two parameterizations of the scalar power spectrum to study this connection.

\subsection{Broken power law}

Firstly, we take a broken power-law function with three different power indices to parameterize the scalar power spectrum with a sharp spike
predicted in various inflationary models \cite{Bullock:1996at,Saito:2008em,Kadota:2015dza,Inomata:2016rbd,Ballesteros:2017fsr,Cicoli:2018asa,Cheng:2018qof,Fu:2019ttf,Bhaumik:2019tvl,Lin:2020goi,Dalianis:2018frf,Tada:2019amh,Fu:2019vqc,Xu:2019bdp,Yi:2020kmq}.
The broken power-law parametrizations with simpler forms were also considered in \cite{Inomata:2016rbd,Inomata:2017okj,Lu:2019sti}.
The form of the parameterization we use is
\begin{equation}\label{E}
y=-ax-b,
\end{equation}
where $y=\log_{10}\mathcal{P}_{\zeta}$ and $x=\log_{10}(k/k_p)$. Parameterizing the power spectrum with four different power-law,
we get
\begin{equation}\label{eq:bpl}
\mathcal{P}_{\zeta}\left(k\right)=\left\{
\begin{array}{rcl}
\mathcal{A}_{\zeta 1}\bar{k}^{n_{1}-1},&&{\tilde{k}\le l/p},\\
\mathcal{A}_{\zeta 2}\bar{k}^{n_{2}-1},&&{l/p<\bar{k}\le 1},\\
\mathcal{A}_{\zeta 2}\bar{k}^{n_{3}-1},&&{1<\bar{k}\le dl/p},\\
\mathcal{A}_{\zeta 1}\bar{k}^{n_{1}-1},&&{\bar{k}>dl/p},
\end{array} \right.
\end{equation}
where the CMB pivotal scale $k_*=0.05$ Mpc$^{-1}$, $l=k_1/k_{*}$, $p=k_p /k_{*}$, $d=k_2/k_1$, $\bar{k}=k/k_{p}$,
$k_1$, $k_p$ and $k_2$ correspond to the scales of the start,
the peak and the end of the spike, respectively,
$\mathcal{A}_{\zeta 1}=\tilde{\mathcal{A}}_{\zeta}p^{n_1-1}$, ${\tilde{\mathcal{A}}_{\zeta}}$ is the amplitude of the power spectrum at CMB pivotal scale
and $\mathcal{A}_{\zeta 2}$ is the amplitude at the peak.
The power spectrum satisfies the CMB constraint at large scales $k<k_1$, so we choose $n_1=0.9649$ and $\tilde{\mathcal{A}}_{\zeta}=2.1\times 10^{-9}$ \cite{Akrami:2018odb}.
For the discussion on the shape of the spike, the power spectrum at $k<k_1$ and $k>k_2$ is not important.
For simplicity, we take the same parameterization of the power spectrum for  $k<k_1$ and $k>k_2$.
To induce observable SIGWs, $\mathcal{A}_{\zeta 2}$ should be in  the order of ${\mathcal{O}\left(10^{-2}\right)}$, here we choose it to be 0.01.
For convenience, instead of the three scales $k_1$, $k_p$ and $k_2$,
we use the three dimensionless parameters $l$, $p$ and $d$.
We fix the peak scale at $p=10^{13}$. Once the power indices $n_2$ and $n_3$ are given,
the parameters $l$ and $d$ can be derived from the following relations
\begin{equation}
\text{ln}l=\frac{\ln(\mathcal{A}_{\zeta 1}/\mathcal{A}_{\zeta 2})+\left(n_{2}-1\right)\text{ln}p}{n_2-n_1},
\end{equation}
and
\begin{equation}
\text{ln}d=\frac{\ln(\mathcal{A}_{\zeta 1}/\mathcal{A}_{\zeta 2})+\left(n_{3}-1\right)\text{ln}p}{n_3-n_1}
-\text{ln}l.
\end{equation}

With the power spectrum and the formulae presented in the previous section,
we can calculate the energy density of SIGWs
and investigate the relationship between the shapes of $\Omega_{GW}$ and $\mathcal{P}_\zeta$ by changing the power indices $n_2$ and $n_3$.

\begin{figure}[htp]
  \centering
  \subfigure{\includegraphics[width=0.45\textwidth]{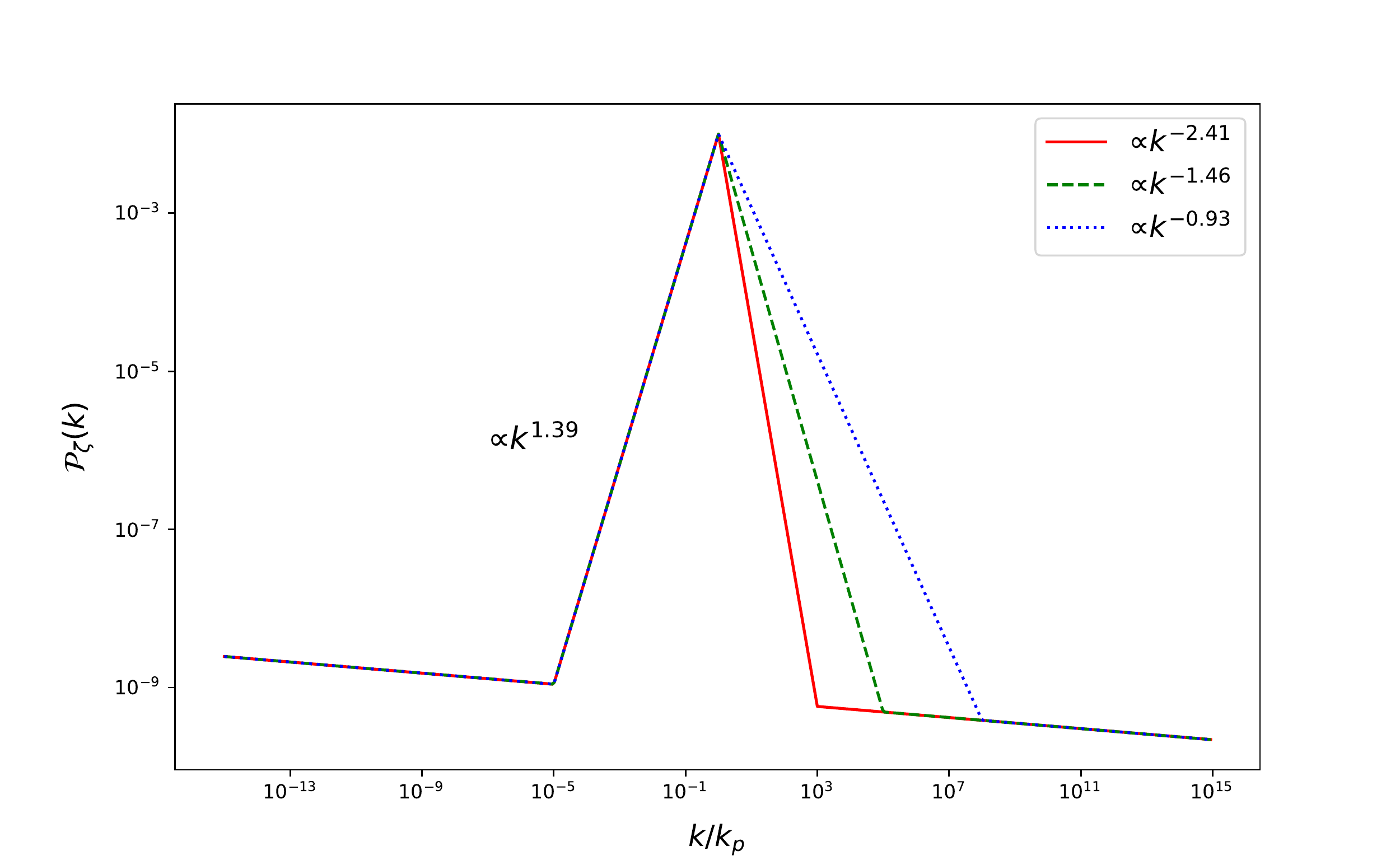}}
      \subfigure{\includegraphics[width=0.45\textwidth]{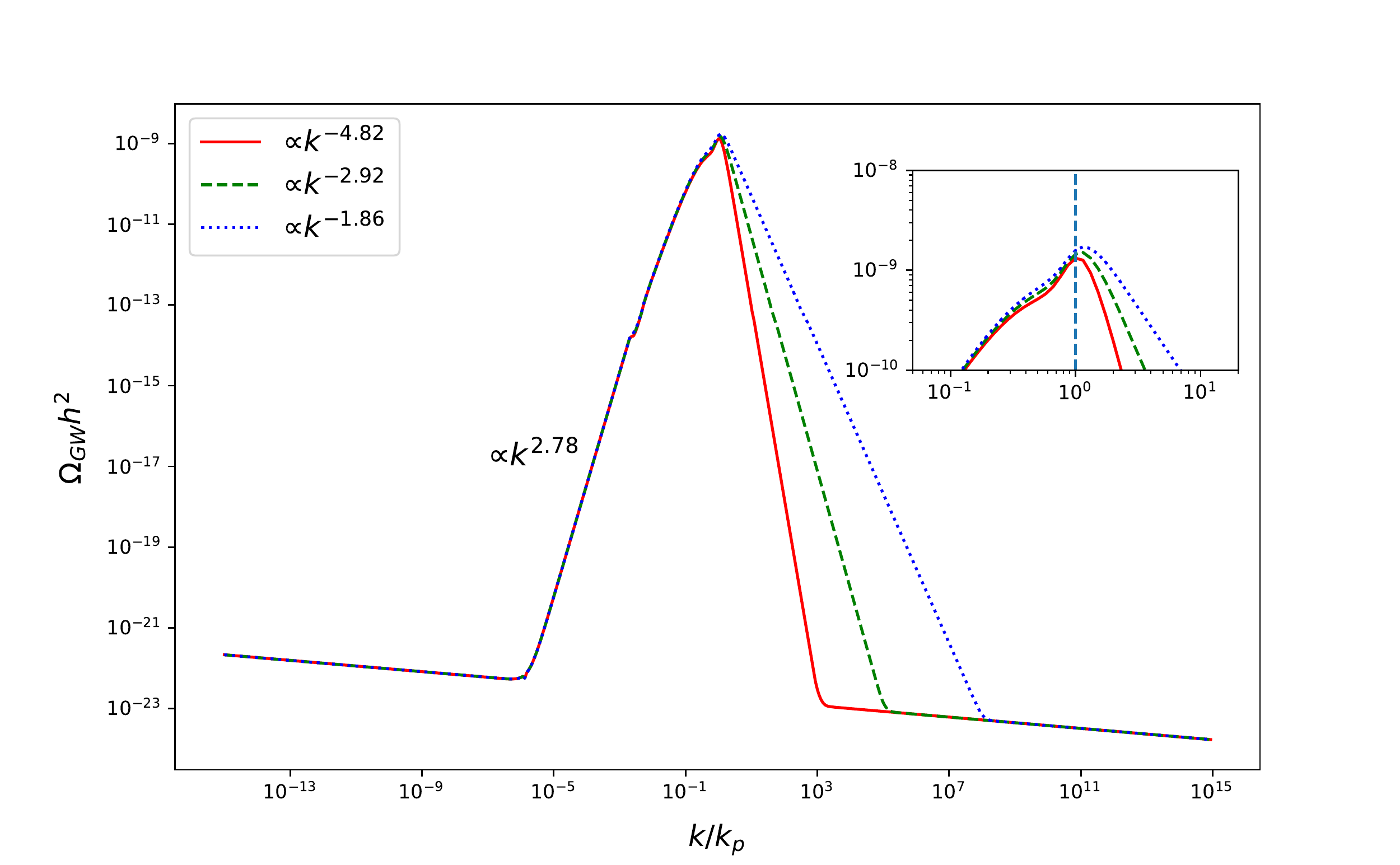}}
  \caption{The scalar power spectrum with the broken power-law parametrization and
the calculated SIGWs with the power spectrum.
The left panel shows the scalar power spectrum.
As for the left side of the peak, the power index $n_{2}-1=1.39$.
As for the right side of the peak, the power indices $n_3-1=-2.41$,
$n_3-1=-1.46$, and $n_3-1=-0.93$ correspond to the solid red, dashed green and dotted blue lines, respectively.
The right panel shows $\Omega_{GW}h^2$, and the insert shows the detailed structure near the peak.
On the right side of the peak, the solid red, dashed green and dotted blue lines can be
approximated by the power-law function
with power indices -4.82, -2.92, and -1.86, respectively.
On the left side of the peak, the lower part is approximated by
the power-law function with the power index 2.78.}
\label{trir}
\end{figure}

First we fix the left side of the spike, i.e., we take $n_2-1=1.39$, and vary
the right side of the spike by taking three different power indices, $n_3-1=-0.93$,
$n_3-1=-1.46$ and $n_3-1=-2.41$.
The results are shown in Fig. \ref{trir}. From Fig. \ref{trir}, we see that
$\Omega_{GW}$ and $\mathcal{P}_{\zeta}$ have a similar broken power-law form for the spike.
For the right side of the spike, $\Omega_{GW}\propto k^{n_{g3}}$ with $n_{g3}\approx2(n_3-1)$ a little away from the peak.
For the left side of the spike,
$\Omega_{GW}$ has the power-law form $ k^{n_{g2}}$ further away from the peak, and there is a small dip near the peak.
Away from the spike, the power index $n_{g2}=2(n_2-1)=2.78$.
Near the peak, the waveform of $\Omega_{GW}$ on the left side of the spike
is affected by the power spectrum on the right side of the spike, and it depends on the power index $n_3$.

\begin{figure}[htp]
  \centering
  \subfigure{\includegraphics[width=0.45\textwidth]{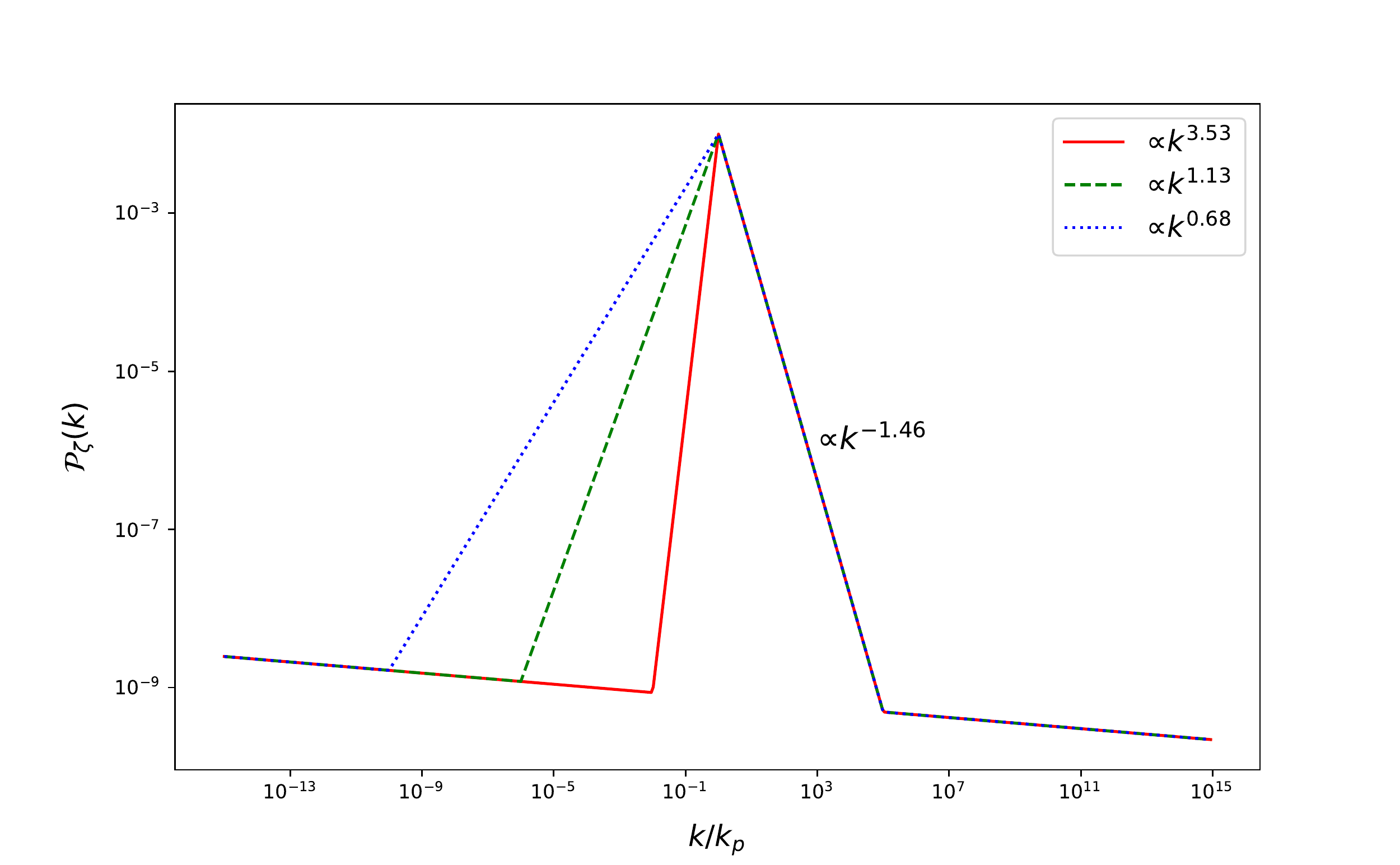}}
  \subfigure{\includegraphics[width=0.45\textwidth]{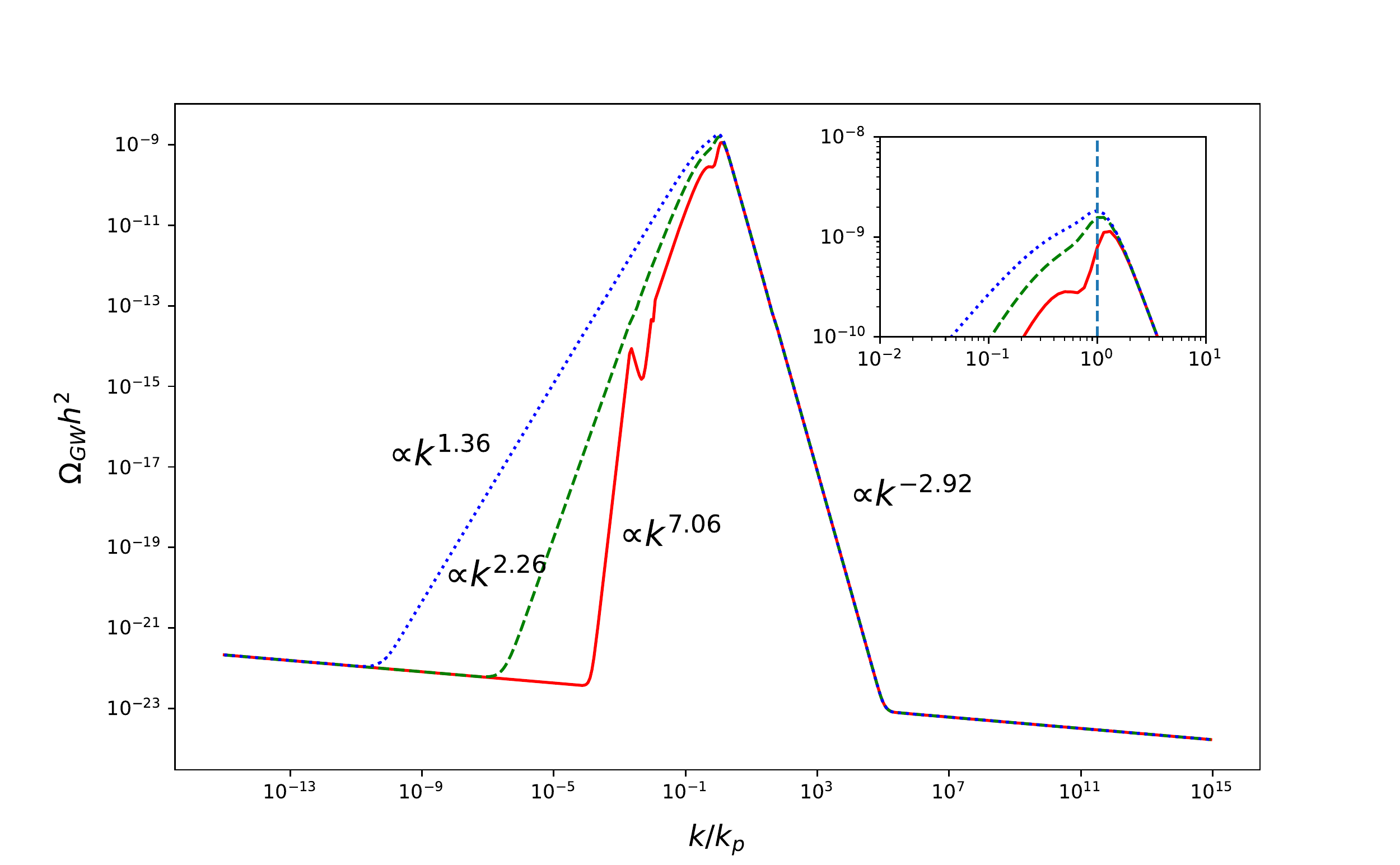}}
  \caption{The scalar power spectrum with the broken power-law parametrization and
the SIGWs generated by the power spectrum.
The left panel shows the scalar power spectrum.
As for the left side of the peak, the power indices
for the solid red, dashed green and dotted blue lines are $n_2-1=3.53$,
$n_2-1=1.13$, and $n_2-1=0.68$, respectively.
As for the right side of the peak, the power index $n_3-1=-1.46$.
The right panel shows $\Omega_{GW}h^2$, and the insert shows the detailed structure near the peak.
On the right side of the peak, $\Omega_{GW}h^2$
can be approximated by the power-law function with the power index -2.92.
On the left side of the peak, the lower part of the solid red, dashed green and dotted blue lines can be
approximated by the power-law function
with power indices 7.06, 2.26, and 1.36, respectively.}\label{tril}
\end{figure}

To see whether the waveform of $\Omega_{GW}$ on the left side of the spike is affected by the
power index $n_2$, we then fix the right side of the spike, i.e. we take $n_3-1=-1.46$, and vary
the left side of the spike by taking three different  power indices, $n_2-1=0.68$,
$n_2-1=1.13$ and $n_2-1=3.53$.
The results are shown in Fig. \ref{tril}. From Fig. \ref{tril}, we see that the
relationship $\Omega_{GW}\propto k^{n_{g3}}$ with $n_{g3}\approx2(n_3-1)=-2.92$ is always satisfied and hardly affected by the change of the power index $n_2$. As for the left side of the spike, when the power index $n_2$ is small, the relationship $n_{g2}\approx2(n_2-1)$ holds except very near the peak.
With the increase of the power index $n_2$, $n_{g2}\approx2(n_2-1)$ holds only when $k$ is further away from the peak $k_p$,
and there is an additional bump if $n_2\gtrsim 2.1$.

For pure power-law power spectrum, the relation $n_g=2(n_s-1)$ can be easily understood from Eq. \eqref{Eph2}. If $\mathcal{P}_\zeta=A k^{n_s-1}$, then
$\mathcal{P}_\zeta(u k)=A u^{n_s-1} k^{n_s-1}$ and $\mathcal{P}_\zeta(v k)=A v^{n_s-1} k^{n_s-1}$, so the $k$ dependence is decoupled from the integral variables $u$ and $v$
and we get $\Omega_{GW}\propto k^{2(n_s-1)}$.
For the case with the broken power-law power spectrum, we can understand
the above results from the integration kernel $I^{2}_{RD}$.
From Eq. \eqref{IRD}, we see that the main contribution to $I^{2}_{RD}$ is from $u\approx v\approx \sqrt{3}/2$ or $\tilde{k}\approx \sqrt{3}k/2$,
this is also called the resonant amplification \cite{Ananda:2006af,Kohri:2018awv} and
it is the reason why the peak in $\Omega_{GW}$ is located at $2k_p/\sqrt{3}$
for the monochromatic power spectrum \cite{Lu:2019sti}.
Away from the peak, ${k\ll k_p}$ or ${k\gg k_p}$, $\mathcal{P}_\zeta(uk)$ and
$\mathcal{P}_\zeta(vk)$ have the same power law form around $u\approx v\approx \sqrt{3}/2$,
so the $k$ dependence is again decoupled from the integral variables $u$ and $v$
and we have $\Omega_{GW}\propto k^{2(n_s-1)}$ as shown
in Figs. \ref{trir} and \ref{tril}.

In the radiation-dominated era, SIGWs are generated almost instantly after the horizon reentry because the scalar source of SIGWs decays quickly \cite{Baumann:2007zm}.
Since large-scale perturbations reenter the horizon later than small-scale ones,
thus the effect of larger scales is small and the contribution to the integral \eqref{Eph1} from modes with $\tilde{k}<k$ is negligible.
Therefore, for a given mode $k$, we need to consider the contribution from smaller scales with $\tilde{k}>k$
in addition to the major contribution around $\tilde{k}\approx \sqrt{3}k/2$ due to the resonant amplification.
On the right side of the spike, when $k\gtrsim 2k_p/\sqrt{3}$,
the major contribution is from the modes with $\tilde{k}>\sqrt{3}k/2>k_p$ and they have the same functional form,
so we have $\Omega_{GW}(k)\sim \mathcal{P}_\zeta^2(k)$ for the frequency dependence and henceforth $n_g=2(n_s-1)$.
This also explains why the shape on the left side of the spike of the power spectrum does not significantly affect the shape of $\Omega_{GW}$ on the right side of the spike.
On the left side of the spike, near the spike, $\Omega_{GW}$ has contributions from both smaller and larger modes with $\tilde{k}<k_p$ and $\tilde{k}>k_p$.
Since the functional form for $\tilde{k}<k_p$ is different from that for $\tilde{k}>k_p$,
so the shape of $\Omega_{GW}$ becomes different from the power spectrum.
Far away from the peak, $k\ll k_p$, the contribution from the right side of the spike is negligible, and again we get the relation $n_g=2(n_s-1)$.

From the above discussion, we learn that the relation $\Omega_{GW}(k)\sim \mathcal{P}_\zeta^2(k)$
for the frequency dependence should hold for more general power spectrum,
we will confirm this with another parameterization of the power spectrum below.

\subsection{Broad power spectrum}

To parameterize the power spectrum with a broad spike, such as the arched shape obtained in some inflationary models \cite{Inomata:2016rbd,Gong:2017qlj,Espinosa:2017sgp,Espinosa:2018eve,Passaglia:2018ixg,Garcia-Bellido:2017mdw,Garcia-Bellido:2017aan,Drees:2019xpp},
we modify the parameterization \eqref{E} by adding a nonlinear term $1/(cx-e)$ with $c>0$ and $e>0$, i.e., we take the parameterization $y=-ax-b+1/(cx-e)$.

With this parameterization, we get
\begin{equation}
\mathcal{P}_{\zeta}\left(k\right)=\left\{
\begin{array}{ccl}
\mathcal{A}_{\zeta 1}\bar{k}^{n_{1}-1},&&{\bar{k}<l/p},\\
\mathcal{A}_{\zeta 2}\bar{k}^{n_{2}-1},&&{l/p\le \bar{k}<1},\\
\mathcal{A}_{\zeta 3}\bar{k}^{-a}10^{1/(c\text{log}_{10}\bar{k}-e)},&&{\bar{k}\ge 1},
\end{array} \right.
\end{equation}
where $n_1=0.9649$, $\mathcal{A}_{\zeta 2}=0.01$, $\mathcal{A}_{\zeta 3}=10^{-b}$, ${l=k_1/k_*}$, $p=k_p/k_*=10^{12}$, $k_*=0.05$ Mpc$^{-1}$, ${\mathcal{A}_{\zeta 1}=\tilde{\mathcal{A}}_{\zeta}p^{n_{1}-1}}$,
${\tilde{\mathcal{A}}_{\zeta}}=2.1\times 10^{-9}$,  ${k_1}$ and ${k_p}$ represent the scales of the start and the peak of the spike, respectively.
In this parameterization, there are five dimensionless parameters $l$, $p$, $a$, $c$ and $e$.
For given parameters $n_2$ and $e$, the scale parameter $l$ and the amplitude parameter $b$ can be obtained from the following relations,
\begin{equation}
\text{ln}l=\frac{\ln(\mathcal{A}_{\zeta 1}/\mathcal{A}_{\zeta 2})+\left(n_{2}-1\right)\text{ln}p}{n_2-n_1},
\end{equation}
\begin{equation}\label{b}
b=-\left(\text{ln}\mathcal{A}_{\zeta 2}+1/e\right).
\end{equation}

\begin{figure}[htp]
  \centering
  \subfigure{\includegraphics[width=0.45\textwidth]{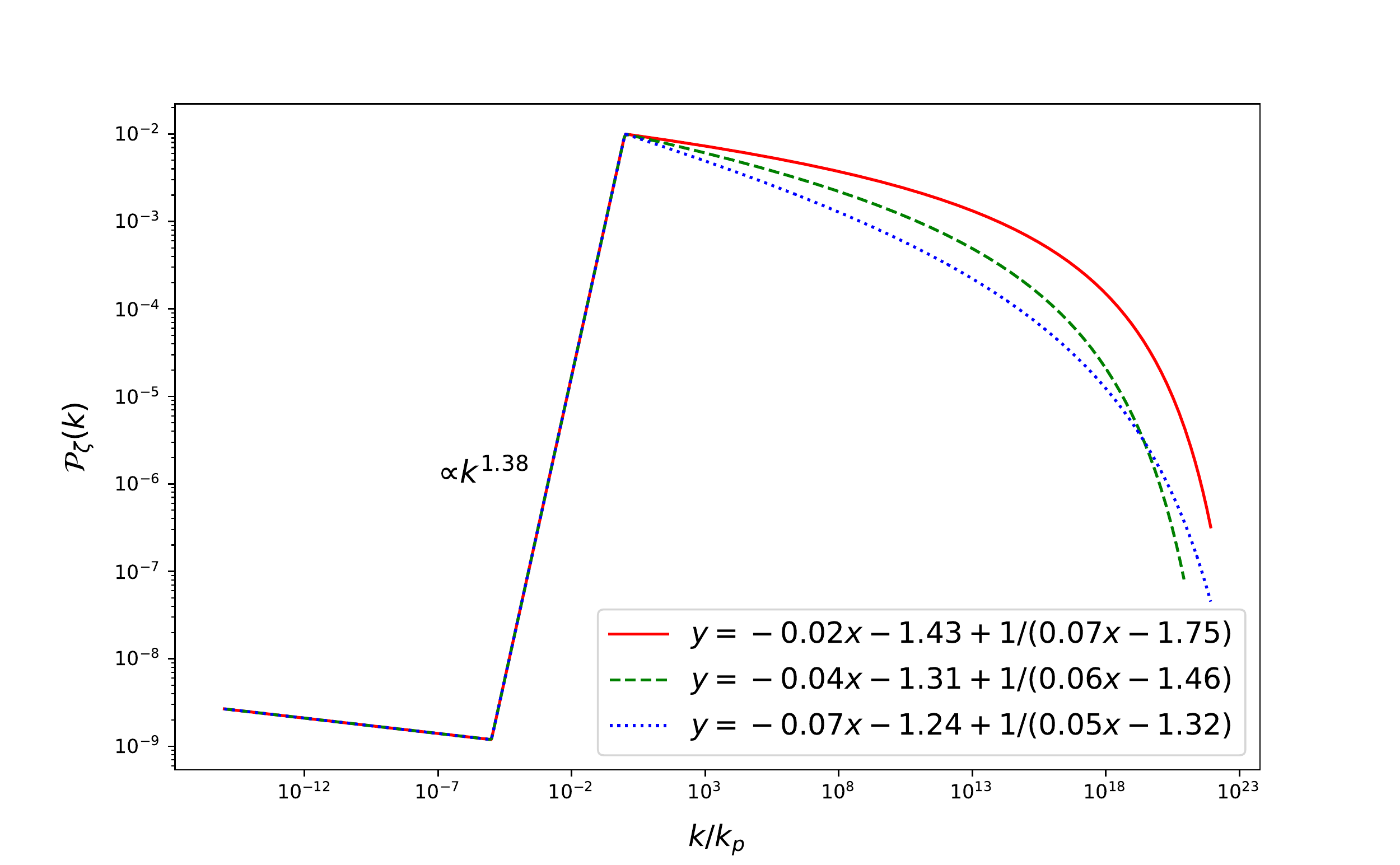}}
     \subfigure{\includegraphics[width=0.45\textwidth]{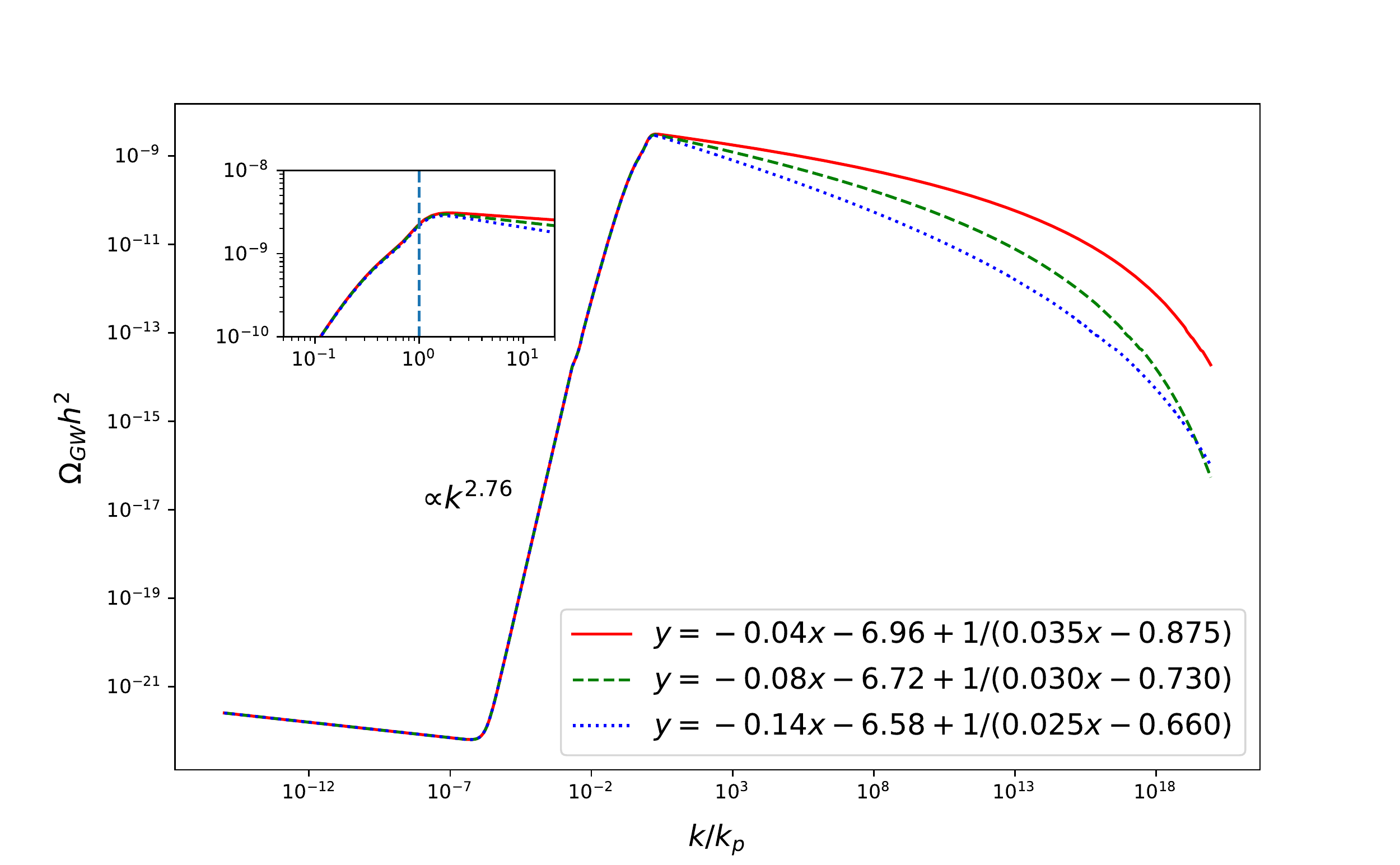}}
  \caption{The scalar power spectrum with a broad spike at small scales and
the calculated SIGWs with the power spectrum.
The left panel shows the scalar power spectrum.
As for the left side of the spike, the power index $n_{2}-1=1.38$.
As for the right side of the spike, the parameters for the solid red, dashed green and dotted blue lines
are $(a,b,c,e)=(0.02,1.43,0.07,1.75)$, $(0.04,1.31,0.06,1.46)$ and $(0.07,1.24,0.05,1.32)$, respectively.
The right panel shows $\Omega_{GW}h^2$ and the insert shows the detailed structure near the peak.
The lower part of the left side of the spike can be approximated
by the power-law function with the power index $n_g=2.76$.
The right side of the spike can be approximated with the
nonlinear function \eqref{omgweq1}. The parameters for
the solid red, dashed green and dotted blue lines are $(a_g,b_g,c_g,e_g)=(0.04,6.96,0.035,0.875)$, $(0.08,6.72,0.03,0.73)$ and $(0.14,6.58,0.025,0.66)$, respectively.}
\label{arcr}
\end{figure}

Again we first fix the left side of the spike by taking the power index $n_2-1=1.38$,
and change the shape of the spike on the right side by choosing different parameters $a$, $c$ and $e$ as shown in the left panel of Fig. \ref{arcr}.
The induced $\Omega_{GW}$ is shown in the right panel of Fig. \ref{arcr}.
We see that the waveform of $\Omega_{GW}$ has a similar arched shape as the power spectrum.
It is also interesting to note that the waveform on the right side of the spike has the same functional form as the power spectrum.
The waveform of $\Omega_{GW}$ can be approximated as
\begin{equation}
\label{omgweq1}
 y_{GW}=\log_{10}\Omega_{GW}(k)=-a_g\log_{10}\bar{k}-b_g+1/(c_g\log_{10}\bar{k}-e_g),
\end{equation}
where $a_g\approx 2a$, $c_g\approx c/2$, $e_g\approx e/2$ and $b_g\approx 2b+4.1$.
For the left side of the spike, similar to the broken power-law case discussed in the previous subsection, $\Omega_{GW}$
has the power-law form $ k^{n_{g2}}$ with $n_{g2}=2(n_2-1)= 2.76$
away from the peak, and there is a small dip near the peak.

\begin{figure}[htp]
  \centering
  \subfigure{\includegraphics[width=0.45\textwidth]{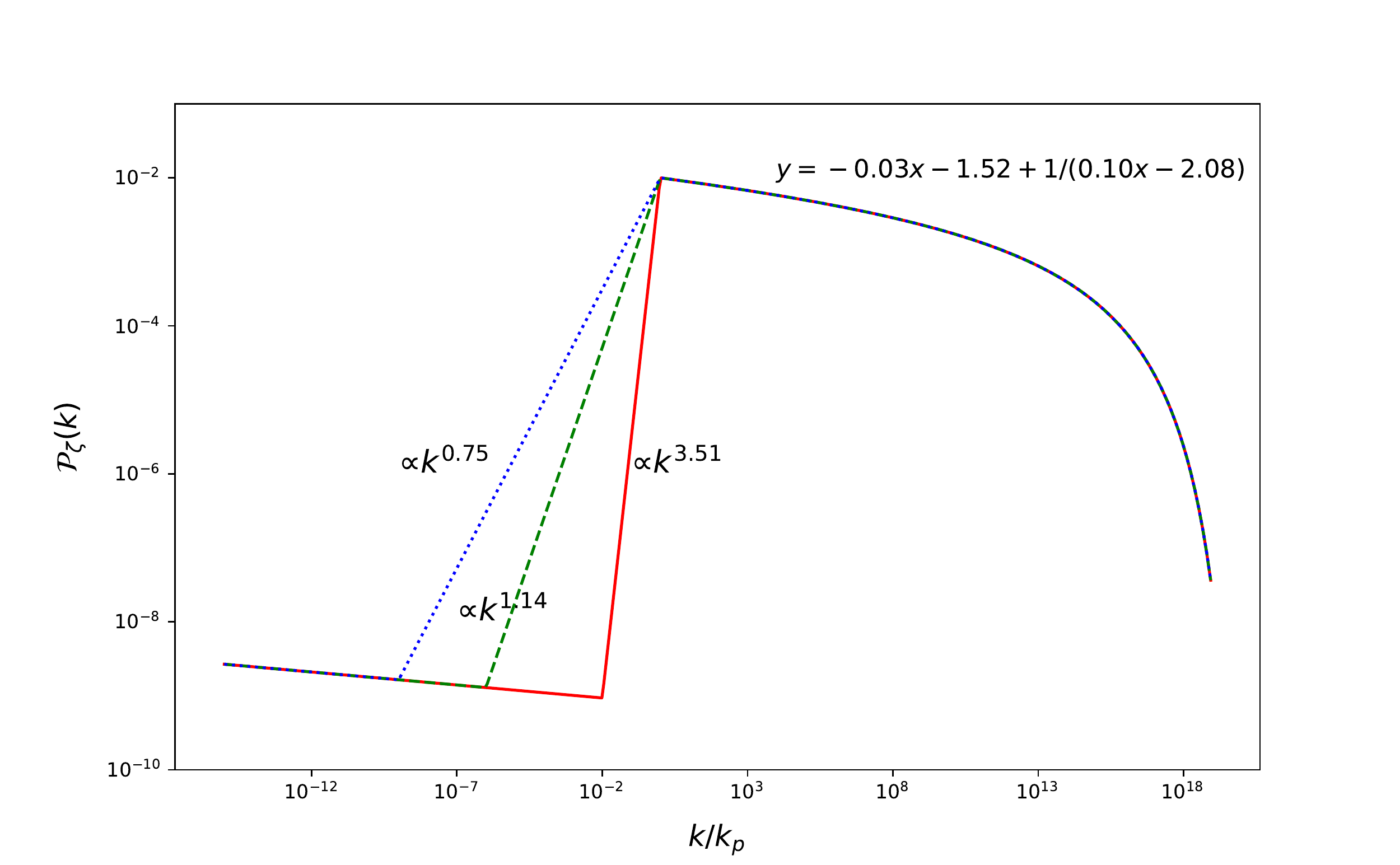}}
  \subfigure{\includegraphics[width=0.45\textwidth]{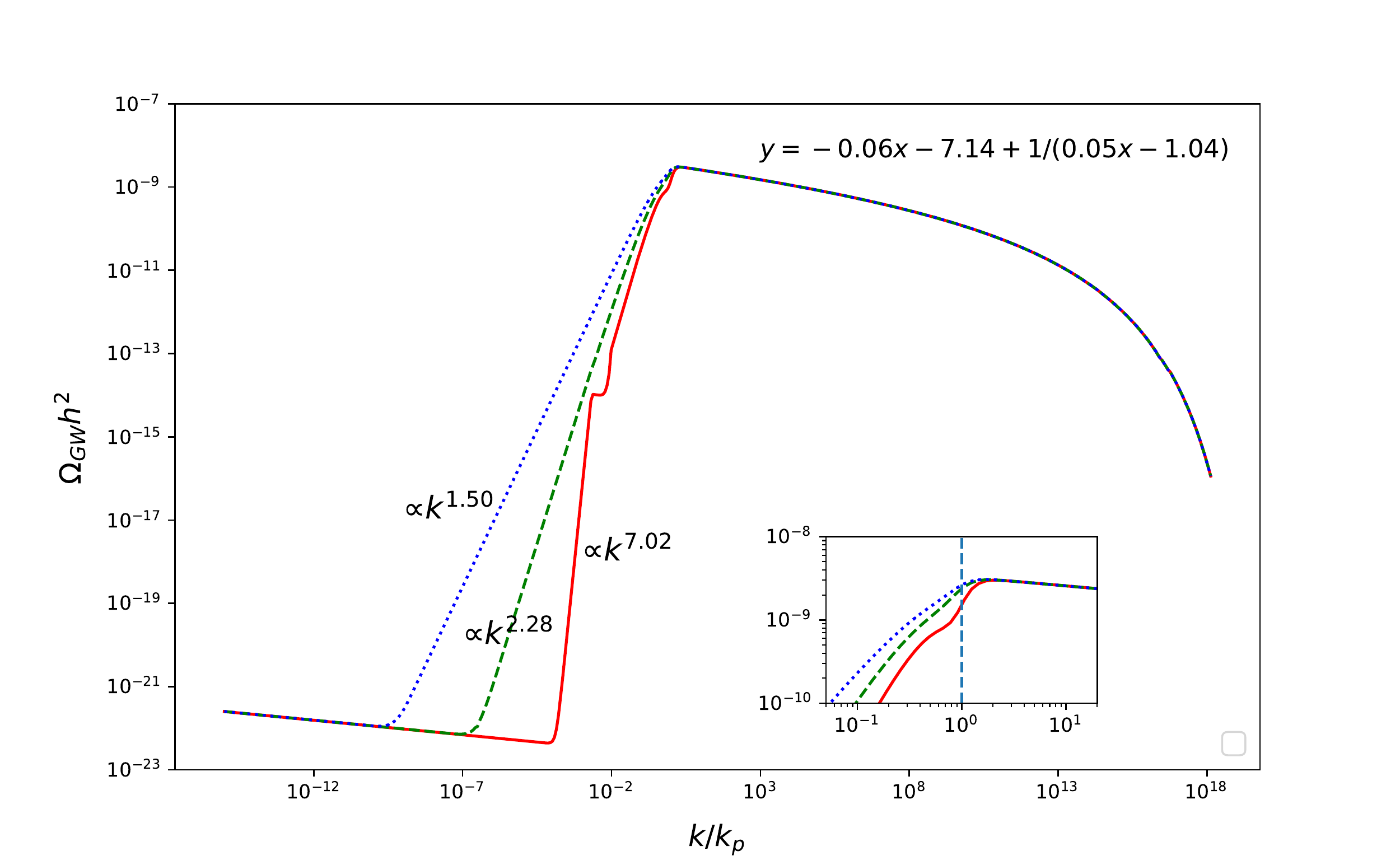}}
  \caption{
  The scalar power spectrum with a broad spike at small scales and
the calculated SIGWs with the power spectrum.
The left panel shows the scalar power spectrum.
The power indices for the solid red, dashed green and dotted blue lines on
the left side of the spike are $n_{2}-1=3.51$, 1.14, and 0.75, respectively.
The parameters for right side of the spike
are $(a,b,c,e)=(0.03,1.52,0.1,2.08)$.
The right panel shows $\Omega_{GW}h^2$, and the insert shows the detailed structure near the peak.
The lower part of the solid red, dashed green and dotted blue lines on the right side of the spike can be approximated
by the power-law function with the power indices $n_g=7.02$, 2.28, and 1.5, respectively.
The right side of the spike can be approximated with the
nonlinear function \eqref{omgweq1} with the parameters  $(a_g,b_g,c_g,e_g)=(0.06,7.14,0.05,1.04)$.}\label{arcl}
\end{figure}

To see whether the power index $n_2$ affects the waveform on the right side of the spike,
we then fix the power spectrum on the right side of the spike by taking the parameters $a=0.03$, $b=1.52$, $c=0.10$, and $e=2.08$,
and vary the power index on the left side of the spike with $n_2-1=0.75$,
$n_2-1=1.14$, and $n_2-1=3.51$.
The results are shown in Fig. \ref{arcl}.
Since the power spectrum on the left side of the spike is power-law form,
so the results of $\Omega_{GW}$ on the left side of the spike are the same as those for the broken power-law case discussed in the previous subsection.
Form Fig. \ref{arcl}, as expected, we see that the power spectrum on the left side of the spike
has no effect on the waveform of the right side of the spike. On the right side of the spike, for the frequency dependence, we have the relation $\Omega_{GW}(k)\propto \mathcal{P}_\zeta^2(k)$ .

As explained in the previous subsection, for the modes a little larger than $k_p$
($k>2k_p/\sqrt{3}$),
we have the relation $\Omega_{GW}(k)\sim \mathcal{P}_\zeta^2(k)$,
so $\Omega_{GW}(k)$ have the same arched shape \eqref{omgweq1} with the parameters $a_{g}\approx 2a$, $c_{g}\approx 1/2c$,
$e_{g}\approx 1/2e$ and $b_{g}\approx 2b+4.1$.
The additional constant 4.1 for the amplitude comes from the current fractional energy density of radiation ${\Omega_{r,0}\simeq \mathcal{O}\left(10^{-4}\right)}$ \cite{Aghanim:2018eyx}. These results confirm that away from the spike
the frequency relation $\Omega_{GW}(k)\sim \mathcal{P}_\zeta^2(k)$ holds
independent of the functional form of the power spectrum
and the waveform of SIGWs has a similar shape to the power spectrum.

\section{Conclusions}
The waveform of the stochastic GW background can be approximated as the broken
power-law form $\Omega_{GW}\approx \Omega_n(f/f_p)^n$
and the power index $n$ tells us the origin of the stochastic GW background.
SIGWs are induced by the scalar perturbation generated during inflation,
so the detection of SIGWs provides us a useful tool to probe the early universe.
To study inflationary models with SIGWs, we need to
know the relation between SIGWs and the power spectrum of the scalar perturbation. In particular, we need to know how the waveform of SIGWs is related
to the shape of the scalar power spectrum.
We propose two parameterizations to approximate the scalar power
spectrum with either a sharp spike or a broad spike at small scales, respectively,
and we find that the waveform of SIGWs has a similar shape to the power spectrum.
Away from the peak of the spike, the frequency relation $\Omega_{GW}(k)\sim \mathcal{P}_\zeta^2(k)$ holds
independent of the functional form of the scalar power spectrum.

Although different modes couple to each other
because of the nonlinear effect,
for the production of SIGWs at a mode $k$,
the main contribution comes from modes with $\tilde{k}\approx \sqrt{3}/2k$
in addition to the non-negligible contributions from the modes with $\tilde{k}>k$.
Therefore, if $k>2k_p/\sqrt{3}$ or $k\ll k_p$,
we have $\Omega_{GW}(k)\sim \mathcal{P}_\zeta^2(k)$ for the frequency dependence.
Around the peak $k\sim k_p$, the waveform of $\Omega_{GW}(k)$ becomes complicated,
and it may have additional bumps due to the nonlinear coupling.
For the unusual power spectrum like the $\delta$
function, the simple square relation does not hold.
In conclusion, with the waveform of SIGWs,
the simple relation $\Omega_{GW}(k)\sim \mathcal{P}_\zeta^2(k)$
is useful to determine the scalar power spectrum and
probe inflationary physics.

\begin{acknowledgments}
This research was supported in part by the National Natural Science Foundation of China under Grant No. 11875136 and the Major Program of the National Natural Science Foundation of China under Grant No. 11690021.
\end{acknowledgments}


\begin{thebibliography}{97}%
\makeatletter
\providecommand \@ifxundefined [1]{%
 \@ifx{#1\undefined}
}%
\providecommand \@ifnum [1]{%
 \ifnum #1\expandafter \@firstoftwo
 \else \expandafter \@secondoftwo
 \fi
}%
\providecommand \@ifx [1]{%
 \ifx #1\expandafter \@firstoftwo
 \else \expandafter \@secondoftwo
 \fi
}%
\providecommand \natexlab [1]{#1}%
\providecommand \enquote  [1]{``#1''}%
\providecommand \bibnamefont  [1]{#1}%
\providecommand \bibfnamefont [1]{#1}%
\providecommand \citenamefont [1]{#1}%
\providecommand \href@noop [0]{\@secondoftwo}%
\providecommand \href [0]{\begingroup \@sanitize@url \@href}%
\providecommand \@href[1]{\@@startlink{#1}\@@href}%
\providecommand \@@href[1]{\endgroup#1\@@endlink}%
\providecommand \@sanitize@url [0]{\catcode `\\12\catcode `\$12\catcode
  `\&12\catcode `\#12\catcode `\^12\catcode `\_12\catcode `\%12\relax}%
\providecommand \@@startlink[1]{}%
\providecommand \@@endlink[0]{}%
\providecommand \url  [0]{\begingroup\@sanitize@url \@url }%
\providecommand \@url [1]{\endgroup\@href {#1}{\urlprefix }}%
\providecommand \urlprefix  [0]{URL }%
\providecommand \Eprint [0]{\href }%
\providecommand \doibase [0]{https://doi.org/}%
\providecommand \selectlanguage [0]{\@gobble}%
\providecommand \bibinfo  [0]{\@secondoftwo}%
\providecommand \bibfield  [0]{\@secondoftwo}%
\providecommand \translation [1]{[#1]}%
\providecommand \BibitemOpen [0]{}%
\providecommand \bibitemStop [0]{}%
\providecommand \bibitemNoStop [0]{.\EOS\space}%
\providecommand \EOS [0]{\spacefactor3000\relax}%
\providecommand \BibitemShut  [1]{\csname bibitem#1\endcsname}%
\let\auto@bib@innerbib\@empty
\bibitem [{\citenamefont {Abbott}\ {\it et~al.}(2016{\natexlab{a}})\citenamefont {Abbott}
  {\it et~al.}}]{Abbott:2016nmj}%
  \BibitemOpen
  \bibfield  {author} {\bibinfo {author} {\bibfnamefont {B.~P.}\ \bibnamefont
  {Abbott}} {\it et~al.} (\bibinfo {collaboration} {Virgo, LIGO
  Scientific}),\ }\bibinfo {title} {{GW151226: Observation of Gravitational
  Waves from a 22-Solar-Mass Binary Black Hole Coalescence}},\ \href
  {https://doi.org/10.1103/PhysRevLett.116.241103} {\bibfield  {journal}
  {\bibinfo  {journal} {Phys. Rev. Lett.}\ }\textbf {\bibinfo {volume} {116}},\
  \bibinfo {pages} {241103} (\bibinfo {year} {2016}{\natexlab{a}})},\ \Eprint
  {https://arxiv.org/abs/1606.04855} {arXiv:1606.04855} \BibitemShut {NoStop}%
\bibitem [{\citenamefont {Abbott}\ \it
  {et~al.}(2016{\natexlab{b}})\citenamefont {Abbott} \it
  {et~al.}}]{Abbott:2016blz}%
  \BibitemOpen
  \bibfield  {author} {\bibinfo {author} {\bibfnamefont {B.~P.}\ \bibnamefont
  {Abbott}} {\it et~al.} (\bibinfo {collaboration} {LIGO Scientific and Virgo
  Collaborations}),\ }\bibinfo {title} {{Observation of Gravitational Waves
  from a Binary Black Hole Merger}},\ \href
  {https://doi.org/10.1103/PhysRevLett.116.061102} {\bibfield  {journal}
  {\bibinfo  {journal} {Phys. Rev. Lett.}\ }\textbf {\bibinfo {volume} {116}},\
  \bibinfo {pages} {061102} (\bibinfo {year} {2016}{\natexlab{b}})},\ \Eprint
  {https://arxiv.org/abs/1602.03837} {arXiv:1602.03837} \BibitemShut {NoStop}%
\bibitem [{\citenamefont {Abbott}\ \it
  {et~al.}(2017{\natexlab{a}})\citenamefont {Abbott} \it
  {et~al.}}]{Abbott:2017gyy}%
  \BibitemOpen
  \bibfield  {author} {\bibinfo {author} {\bibfnamefont {B.~P.}\ \bibnamefont
  {Abbott}} {\it et~al.} (\bibinfo {collaboration} {Virgo, LIGO
  Scientific}),\ }\bibinfo {title} {{GW170608: Observation of a 19-solar-mass
  Binary Black Hole Coalescence}},\ \href
  {https://doi.org/10.3847/2041-8213/aa9f0c} {\bibfield  {journal} {\bibinfo
  {journal} {Astrophys. J.}\ }\textbf {\bibinfo {volume} {851}},\ \bibinfo
  {pages} {L35} (\bibinfo {year} {2017}{\natexlab{a}})},\ \Eprint
  {https://arxiv.org/abs/1711.05578} {arXiv:1711.05578} \BibitemShut {NoStop}%
\bibitem [{\citenamefont {Abbott}\ \it
  {et~al.}(2017{\natexlab{b}})\citenamefont {Abbott} \it
  {et~al.}}]{TheLIGOScientific:2017qsa}%
  \BibitemOpen
  \bibfield  {author} {\bibinfo {author} {\bibfnamefont {B.~P.}\ \bibnamefont
  {Abbott}} {\it et~al.} (\bibinfo {collaboration} {Virgo, LIGO
  Scientific}),\ }\bibinfo {title} {{GW170817: Observation of Gravitational
  Waves from a Binary Neutron Star Inspiral}},\ \href
  {https://doi.org/10.1103/PhysRevLett.119.161101} {\bibfield  {journal}
  {\bibinfo  {journal} {Phys. Rev. Lett.}\ }\textbf {\bibinfo {volume} {119}},\
  \bibinfo {pages} {161101} (\bibinfo {year} {2017}{\natexlab{b}})},\ \Eprint
  {https://arxiv.org/abs/1710.05832} {arXiv:1710.05832} \BibitemShut {NoStop}%
\bibitem [{\citenamefont {Abbott}\ \it
  {et~al.}(2017{\natexlab{c}})\citenamefont {Abbott} \it
  {et~al.}}]{Abbott:2017oio}%
  \BibitemOpen
  \bibfield  {author} {\bibinfo {author} {\bibfnamefont {B.~P.}\ \bibnamefont
  {Abbott}} {\it et~al.} (\bibinfo {collaboration} {Virgo, LIGO
  Scientific}),\ }\bibinfo {title} {{GW170814: A Three-Detector Observation of
  Gravitational Waves from a Binary Black Hole Coalescence}},\ \href
  {https://doi.org/10.1103/PhysRevLett.119.141101} {\bibfield  {journal}
  {\bibinfo  {journal} {Phys. Rev. Lett.}\ }\textbf {\bibinfo {volume} {119}},\
  \bibinfo {pages} {141101} (\bibinfo {year} {2017}{\natexlab{c}})},\ \Eprint
  {https://arxiv.org/abs/1709.09660} {arXiv:1709.09660} \BibitemShut {NoStop}%
\bibitem [{\citenamefont {Abbott}\ \it
  {et~al.}(2017{\natexlab{d}})\citenamefont {Abbott} \it
  {et~al.}}]{Abbott:2017vtc}%
  \BibitemOpen
  \bibfield  {author} {\bibinfo {author} {\bibfnamefont {B.~P.}\ \bibnamefont
  {Abbott}} {\it et~al.} (\bibinfo {collaboration} {VIRGO, LIGO
  Scientific}),\ }\bibinfo {title} {{GW170104: Observation of a 50-Solar-Mass
  Binary Black Hole Coalescence at Redshift 0.2}},\ \href
  {https://doi.org/10.1103/PhysRevLett.118.221101} {\bibfield  {journal}
  {\bibinfo  {journal} {Phys. Rev. Lett.}\ }\textbf {\bibinfo {volume} {118}},\
  \bibinfo {pages} {221101} (\bibinfo {year} {2017}{\natexlab{d}})},\ \Eprint
  {https://arxiv.org/abs/1706.01812} {arXiv:1706.01812} \BibitemShut {NoStop}%
\bibitem [{\citenamefont {Abbott}\ {\it et~al.}(2019)\citenamefont {Abbott}
  {\it et~al.}}]{LIGOScientific:2018mvr}%
  \BibitemOpen
  \bibfield  {author} {\bibinfo {author} {\bibfnamefont {B.}~\bibnamefont
  {Abbott}} {\it et~al.} (\bibinfo {collaboration} {LIGO Scientific,
  Virgo}),\ }\bibinfo {title} {{GWTC-1: A Gravitational-Wave Transient Catalog
  of Compact Binary Mergers Observed by LIGO and Virgo during the First and
  Second Observing Runs}},\ \href {https://doi.org/10.1103/PhysRevX.9.031040}
  {\bibfield  {journal} {\bibinfo  {journal} {Phys. Rev. X}\ }\textbf {\bibinfo
  {volume} {9}},\ \bibinfo {pages} {031040} (\bibinfo {year} {2019})},\ \Eprint
  {https://arxiv.org/abs/1811.12907} {arXiv:1811.12907} \BibitemShut {NoStop}%
\bibitem [{\citenamefont {Abbott}\ \it
  {et~al.}(2020{\natexlab{a}})\citenamefont {Abbott} \it
  {et~al.}}]{Abbott:2020khf}%
  \BibitemOpen
  \bibfield  {author} {\bibinfo {author} {\bibfnamefont {R.}~\bibnamefont
  {Abbott}} {\it et~al.} (\bibinfo {collaboration} {LIGO Scientific,
  Virgo}),\ }\bibinfo {title} {{GW190814: Gravitational Waves from the
  Coalescence of a 23 Solar Mass Black Hole with a 2.6 Solar Mass Compact
  Object}},\ \href {https://doi.org/10.3847/2041-8213/ab960f} {\bibfield
  {journal} {\bibinfo  {journal} {Astrophys. J.}\ }\textbf {\bibinfo {volume}
  {896}},\ \bibinfo {pages} {L44} (\bibinfo {year} {2020}{\natexlab{a}})},\
  \Eprint {https://arxiv.org/abs/2006.12611} {arXiv:2006.12611} \BibitemShut
  {NoStop}%
\bibitem [{\citenamefont {Abbott}\ \it
  {et~al.}(2020{\natexlab{b}})\citenamefont {Abbott} \it
  {et~al.}}]{Abbott:2020uma}%
  \BibitemOpen
  \bibfield  {author} {\bibinfo {author} {\bibfnamefont {B.}~\bibnamefont
  {Abbott}} {\it et~al.} (\bibinfo {collaboration} {LIGO Scientific,
  Virgo}),\ }\bibinfo {title} {{GW190425: Observation of a Compact Binary
  Coalescence with Total Mass $\sim 3.4 M_{\odot}$}},\ \href
  {https://doi.org/10.3847/2041-8213/ab75f5} {\bibfield  {journal} {\bibinfo
  {journal} {Astrophys. J. Lett.}\ }\textbf {\bibinfo {volume} {892}},\
  \bibinfo {pages} {L3} (\bibinfo {year} {2020}{\natexlab{b}})},\ \Eprint
  {https://arxiv.org/abs/2001.01761} {arXiv:2001.01761} \BibitemShut {NoStop}%
\bibitem [{\citenamefont {Abbott}\ \it
  {et~al.}(2020{\natexlab{c}})\citenamefont {Abbott} \it
  {et~al.}}]{LIGOScientific:2020stg}%
  \BibitemOpen
  \bibfield  {author} {\bibinfo {author} {\bibfnamefont {R.}~\bibnamefont
  {Abbott}} {\it et~al.} (\bibinfo {collaboration} {LIGO Scientific,
  Virgo}),\ }\bibinfo {title} {{GW190412: Observation of a Binary-Black-Hole
  Coalescence with Asymmetric Masses}},\ \Eprint
  {https://arxiv.org/abs/2004.08342} {arXiv:2004.08342} \BibitemShut {NoStop}%
\bibitem [{\citenamefont {Carr}\ and\ \citenamefont
  {Hawking}(1974)}]{Carr:1974nx}%
  \BibitemOpen
  \bibfield  {author} {\bibinfo {author} {\bibfnamefont {B.~J.}\ \bibnamefont
  {Carr}}\ and\ \bibinfo {author} {\bibfnamefont {S.~W.}\ \bibnamefont
  {Hawking}},\ }\bibinfo {title} {{Black holes in the early Universe}},\ \href
  {https://doi.org/10.1093/mnras/168.2.399} {\bibfield  {journal} {\bibinfo
  {journal} {Mon. Not. Roy. Astron. Soc.}\ }\textbf {\bibinfo {volume} {168}},\
  \bibinfo {pages} {399} (\bibinfo {year} {1974})}\BibitemShut {NoStop}%
\bibitem [{\citenamefont {Hawking}(1971)}]{Hawking:1971ei}%
  \BibitemOpen
  \bibfield  {author} {\bibinfo {author} {\bibfnamefont {S.}~\bibnamefont
  {Hawking}},\ }\bibinfo {title} {{Gravitationally collapsed objects of very
  low mass}},\ \href@noop {} {\bibfield  {journal} {\bibinfo  {journal} {Mon.
  Not. Roy. Astron. Soc.}\ }\textbf {\bibinfo {volume} {152}},\ \bibinfo
  {pages} {75} (\bibinfo {year} {1971})}\BibitemShut {NoStop}%
\bibitem [{\citenamefont {Ivanov}\ {\it et~al.}(1994)\citenamefont {Ivanov},
  \citenamefont {Naselsky},\ and\ \citenamefont {Novikov}}]{Ivanov:1994pa}%
  \BibitemOpen
  \bibfield  {author} {\bibinfo {author} {\bibfnamefont {P.}~\bibnamefont
  {Ivanov}}, \bibinfo {author} {\bibfnamefont {P.}~\bibnamefont {Naselsky}},\
  and\ \bibinfo {author} {\bibfnamefont {I.}~\bibnamefont {Novikov}},\
  }\bibinfo {title} {{Inflation and primordial black holes as dark matter}},\
  \href {https://doi.org/10.1103/PhysRevD.50.7173} {\bibfield  {journal}
  {\bibinfo  {journal} {Phys. Rev. D}\ }\textbf {\bibinfo {volume} {50}},\
  \bibinfo {pages} {7173} (\bibinfo {year} {1994})}\BibitemShut {NoStop}%
\bibitem [{\citenamefont {Frampton}\ {\it et~al.}(2010)\citenamefont
  {Frampton}, \citenamefont {Kawasaki}, \citenamefont {Takahashi},\ and\
  \citenamefont {Yanagida}}]{Frampton:2010sw}%
  \BibitemOpen
  \bibfield  {author} {\bibinfo {author} {\bibfnamefont {P.~H.}\ \bibnamefont
  {Frampton}}, \bibinfo {author} {\bibfnamefont {M.}~\bibnamefont {Kawasaki}},
  \bibinfo {author} {\bibfnamefont {F.}~\bibnamefont {Takahashi}},\ and\
  \bibinfo {author} {\bibfnamefont {T.~T.}\ \bibnamefont {Yanagida}},\
  }\bibinfo {title} {{Primordial Black Holes as All Dark Matter}},\ \href
  {https://doi.org/10.1088/1475-7516/2010/04/023} {J. Cosmol. Astropart. Phys.\
  \bibinfo {volume} {1004}\bibfield  {year} {\bibinfo  {year} { (\textbf
  {2010})}\ }\bibfield  {pages} {\bibinfo  {pages} {023}},\ }\Eprint
  {https://arxiv.org/abs/1001.2308} {arXiv:1001.2308} \BibitemShut {NoStop}%
\bibitem [{\citenamefont {Belotsky}\ {\it et~al.}(2014)\citenamefont
  {Belotsky}, \citenamefont {Dmitriev}, \citenamefont {Esipova}, \citenamefont
  {Gani}, \citenamefont {Grobov}, \citenamefont {Khlopov}, \citenamefont
  {Kirillov}, \citenamefont {Rubin},\ and\ \citenamefont
  {Svadkovsky}}]{Belotsky:2014kca}%
  \BibitemOpen
  \bibfield  {author} {\bibinfo {author} {\bibfnamefont {K.~M.}\ \bibnamefont
  {Belotsky}}, \bibinfo {author} {\bibfnamefont {A.~D.}\ \bibnamefont
  {Dmitriev}}, \bibinfo {author} {\bibfnamefont {E.~A.}\ \bibnamefont
  {Esipova}}, \bibinfo {author} {\bibfnamefont {V.~A.}\ \bibnamefont {Gani}},
  \bibinfo {author} {\bibfnamefont {A.~V.}\ \bibnamefont {Grobov}}, \bibinfo
  {author} {\bibfnamefont {M.~{\relax Yu}.}\ \bibnamefont {Khlopov}}, \bibinfo
  {author} {\bibfnamefont {A.~A.}\ \bibnamefont {Kirillov}}, \bibinfo {author}
  {\bibfnamefont {S.~G.}\ \bibnamefont {Rubin}},\ and\ \bibinfo {author}
  {\bibfnamefont {I.~V.}\ \bibnamefont {Svadkovsky}},\ }\bibinfo {title}
  {{Signatures of primordial black hole dark matter}},\ \href
  {https://doi.org/10.1142/S0217732314400057} {\bibfield  {journal} {\bibinfo
  {journal} {Mod. Phys. Lett. A}\ }\textbf {\bibinfo {volume} {29}},\ \bibinfo
  {pages} {1440005} (\bibinfo {year} {2014})},\ \Eprint
  {https://arxiv.org/abs/1410.0203} {arXiv:1410.0203} \BibitemShut {NoStop}%
\bibitem [{\citenamefont {Khlopov}\ {\it et~al.}(2005)\citenamefont
  {Khlopov}, \citenamefont {Rubin},\ and\ \citenamefont
  {Sakharov}}]{Khlopov:2004sc}%
  \BibitemOpen
  \bibfield  {author} {\bibinfo {author} {\bibfnamefont {M.~{\relax Yu}.}\
  \bibnamefont {Khlopov}}, \bibinfo {author} {\bibfnamefont {S.~G.}\
  \bibnamefont {Rubin}},\ and\ \bibinfo {author} {\bibfnamefont {A.~S.}\
  \bibnamefont {Sakharov}},\ }\bibinfo {title} {{Primordial structure of
  massive black hole clusters}},\ \href
  {https://doi.org/10.1016/j.astropartphys.2004.12.002} {\bibfield  {journal}
  {\bibinfo  {journal} {Astropart. Phys.}\ }\textbf {\bibinfo {volume} {23}},\
  \bibinfo {pages} {265} (\bibinfo {year} {2005})},\ \Eprint
  {https://arxiv.org/abs/astro-ph/0401532} {arXiv:astro-ph/0401532}
  \BibitemShut {NoStop}%
\bibitem [{\citenamefont {Clesse}\ and\ \citenamefont
  {Garc\'{i}a-Bellido}(2015)}]{Clesse:2015wea}%
  \BibitemOpen
  \bibfield  {author} {\bibinfo {author} {\bibfnamefont {S.}~\bibnamefont
  {Clesse}}\ and\ \bibinfo {author} {\bibfnamefont {J.}~\bibnamefont
  {Garc\'{i}a-Bellido}},\ }\bibinfo {title} {{Massive Primordial Black Holes
  from Hybrid Inflation as Dark Matter and the seeds of Galaxies}},\ \href
  {https://doi.org/10.1103/PhysRevD.92.023524} {\bibfield  {journal} {\bibinfo
  {journal} {Phys. Rev. D}\ }\textbf {\bibinfo {volume} {92}},\ \bibinfo
  {pages} {023524} (\bibinfo {year} {2015})},\ \Eprint
  {https://arxiv.org/abs/1501.07565} {arXiv:1501.07565} \BibitemShut {NoStop}%
\bibitem [{\citenamefont {Carr}\ {\it et~al.}(2016)\citenamefont {Carr},
  \citenamefont {Kuhnel},\ and\ \citenamefont {Sandstad}}]{Carr:2016drx}%
  \BibitemOpen
  \bibfield  {author} {\bibinfo {author} {\bibfnamefont {B.}~\bibnamefont
  {Carr}}, \bibinfo {author} {\bibfnamefont {F.}~\bibnamefont {Kuhnel}},\ and\
  \bibinfo {author} {\bibfnamefont {M.}~\bibnamefont {Sandstad}},\ }\bibinfo
  {title} {{Primordial Black Holes as Dark Matter}},\ \href
  {https://doi.org/10.1103/PhysRevD.94.083504} {\bibfield  {journal} {\bibinfo
  {journal} {Phys. Rev. D}\ }\textbf {\bibinfo {volume} {94}},\ \bibinfo
  {pages} {083504} (\bibinfo {year} {2016})},\ \Eprint
  {https://arxiv.org/abs/1607.06077} {arXiv:1607.06077} \BibitemShut {NoStop}%
\bibitem [{\citenamefont {Pi}\ {\it et~al.}(2018)\citenamefont {Pi},
  \citenamefont {Zhang}, \citenamefont {Huang},\ and\ \citenamefont
  {Sasaki}}]{Pi:2017gih}%
  \BibitemOpen
  \bibfield  {author} {\bibinfo {author} {\bibfnamefont {S.}~\bibnamefont
  {Pi}}, \bibinfo {author} {\bibfnamefont {Y.-l.}\ \bibnamefont {Zhang}},
  \bibinfo {author} {\bibfnamefont {Q.-G.}\ \bibnamefont {Huang}},\ and\
  \bibinfo {author} {\bibfnamefont {M.}~\bibnamefont {Sasaki}},\ }\bibinfo
  {title} {{Scalaron from $R^2$-gravity as a heavy field}},\ \href
  {https://doi.org/10.1088/1475-7516/2018/05/042} {J. Cosmol. Astropart. Phys.\
  \bibinfo {volume} {1805}\bibfield  {year} {\bibinfo  {year} { (\textbf
  {2018})}\ }\bibfield  {pages} {\bibinfo  {pages} {042}},\ }\Eprint
  {https://arxiv.org/abs/1712.09896} {arXiv:1712.09896} \BibitemShut {NoStop}%
\bibitem [{\citenamefont {Inomata}\ \it
  {et~al.}(2017{\natexlab{a}})\citenamefont {Inomata}, \citenamefont
  {Kawasaki}, \citenamefont {Mukaida}, \citenamefont {Tada},\ and\
  \citenamefont {Yanagida}}]{Inomata:2017okj}%
  \BibitemOpen
  \bibfield  {author} {\bibinfo {author} {\bibfnamefont {K.}~\bibnamefont
  {Inomata}}, \bibinfo {author} {\bibfnamefont {M.}~\bibnamefont {Kawasaki}},
  \bibinfo {author} {\bibfnamefont {K.}~\bibnamefont {Mukaida}}, \bibinfo
  {author} {\bibfnamefont {Y.}~\bibnamefont {Tada}},\ and\ \bibinfo {author}
  {\bibfnamefont {T.~T.}\ \bibnamefont {Yanagida}},\ }\bibinfo {title}
  {{Inflationary Primordial Black Holes as All Dark Matter}},\ \href
  {https://doi.org/10.1103/PhysRevD.96.043504} {\bibfield  {journal} {\bibinfo
  {journal} {Phys. Rev. D}\ }\textbf {\bibinfo {volume} {96}},\ \bibinfo
  {pages} {043504} (\bibinfo {year} {2017}{\natexlab{a}})},\ \Eprint
  {https://arxiv.org/abs/1701.02544} {arXiv:1701.02544} \BibitemShut {NoStop}%
\bibitem [{\citenamefont {Garc\'{i}a-Bellido}(2017)}]{Garcia-Bellido:2017fdg}%
  \BibitemOpen
  \bibfield  {author} {\bibinfo {author} {\bibfnamefont {J.}~\bibnamefont
  {Garc\'{i}a-Bellido}},\ }\bibinfo {title} {{Massive Primordial Black Holes
  as Dark Matter and their detection with Gravitational Waves}},\ \href
  {https://doi.org/10.1088/1742-6596/840/1/012032} {\bibfield  {journal}
  {\bibinfo  {journal} {J. Phys. Conf. Ser.}\ }\textbf {\bibinfo {volume}
  {840}},\ \bibinfo {pages} {012032} (\bibinfo {year} {2017})},\ \Eprint
  {https://arxiv.org/abs/1702.08275} {arXiv:1702.08275} \BibitemShut {NoStop}%
\bibitem [{\citenamefont {Kovetz}(2017)}]{Kovetz:2017rvv}%
  \BibitemOpen
  \bibfield  {author} {\bibinfo {author} {\bibfnamefont {E.~D.}\ \bibnamefont
  {Kovetz}},\ }\bibinfo {title} {{Probing Primordial-Black-Hole Dark Matter
  with Gravitational Waves}},\ \href
  {https://doi.org/10.1103/PhysRevLett.119.131301} {\bibfield  {journal}
  {\bibinfo  {journal} {Phys. Rev. Lett.}\ }\textbf {\bibinfo {volume} {119}},\
  \bibinfo {pages} {131301} (\bibinfo {year} {2017})},\ \Eprint
  {https://arxiv.org/abs/1705.09182} {arXiv:1705.09182} \BibitemShut {NoStop}%
\bibitem [{\citenamefont {Bird}\ {\it et~al.}(2016)\citenamefont {Bird},
  \citenamefont {Cholis}, \citenamefont {Mu\~{n}oz}, \citenamefont
  {Ali-Haimoud}, \citenamefont {Kamionkowski}, \citenamefont {Kovetz},
  \citenamefont {Raccanelli},\ and\ \citenamefont {Riess}}]{Bird:2016dcv}%
  \BibitemOpen
  \bibfield  {author} {\bibinfo {author} {\bibfnamefont {S.}~\bibnamefont
  {Bird}}, \bibinfo {author} {\bibfnamefont {I.}~\bibnamefont {Cholis}},
  \bibinfo {author} {\bibfnamefont {J.~B.}\ \bibnamefont {Mu\~{n}oz}}, \bibinfo
  {author} {\bibfnamefont {Y.}~\bibnamefont {Ali-Haimoud}}, \bibinfo {author}
  {\bibfnamefont {M.}~\bibnamefont {Kamionkowski}}, \bibinfo {author}
  {\bibfnamefont {E.~D.}\ \bibnamefont {Kovetz}}, \bibinfo {author}
  {\bibfnamefont {A.}~\bibnamefont {Raccanelli}},\ and\ \bibinfo {author}
  {\bibfnamefont {A.~G.}\ \bibnamefont {Riess}},\ }\bibinfo {title} {{Did LIGO
  detect dark matter?}},\ \href
  {https://doi.org/10.1103/PhysRevLett.116.201301} {\bibfield  {journal}
  {\bibinfo  {journal} {Phys. Rev. Lett.}\ }\textbf {\bibinfo {volume} {116}},\
  \bibinfo {pages} {201301} (\bibinfo {year} {2016})},\ \Eprint
  {https://arxiv.org/abs/1603.00464} {arXiv:1603.00464} \BibitemShut {NoStop}%
\bibitem [{\citenamefont {Sasaki}\ {\it et~al.}(2016)\citenamefont {Sasaki},
  \citenamefont {Suyama}, \citenamefont {Tanaka},\ and\ \citenamefont
  {Yokoyama}}]{Sasaki:2016jop}%
  \BibitemOpen
  \bibfield  {author} {\bibinfo {author} {\bibfnamefont {M.}~\bibnamefont
  {Sasaki}}, \bibinfo {author} {\bibfnamefont {T.}~\bibnamefont {Suyama}},
  \bibinfo {author} {\bibfnamefont {T.}~\bibnamefont {Tanaka}},\ and\ \bibinfo
  {author} {\bibfnamefont {S.}~\bibnamefont {Yokoyama}},\ }\bibinfo {title}
  {{Primordial Black Hole Scenario for the Gravitational-Wave Event
  GW150914}},\ \href {https://doi.org/10.1103/PhysRevLett.121.059901,
  10.1103/PhysRevLett.117.061101} {\bibfield  {journal} {\bibinfo  {journal}
  {Phys. Rev. Lett.}\ }\textbf {\bibinfo {volume} {117}},\ \bibinfo {pages}
  {061101} (\bibinfo {year} {2016})},\ \bibinfo {note} {[erratum: Phys. Rev.
  Lett.121,no.5,059901(2018)]},\ \Eprint {https://arxiv.org/abs/1603.08338}
  {arXiv:1603.08338} \BibitemShut {NoStop}%
\bibitem [{\citenamefont {Scholtz}\ and\ \citenamefont
  {Unwin}(2020)}]{Scholtz:2019csj}%
  \BibitemOpen
  \bibfield  {author} {\bibinfo {author} {\bibfnamefont {J.}~\bibnamefont
  {Scholtz}}\ and\ \bibinfo {author} {\bibfnamefont {J.}~\bibnamefont
  {Unwin}},\ }\bibinfo {title} {{What if Planet 9 is a Primordial Black
  Hole?}},\ \href {https://doi.org/10.1103/PhysRevLett.125.051103} {\bibfield
  {journal} {\bibinfo  {journal} {Phys. Rev. Lett.}\ }\textbf {\bibinfo
  {volume} {125}},\ \bibinfo {pages} {051103} (\bibinfo {year} {2020})},\
  \Eprint {https://arxiv.org/abs/1909.11090} {arXiv:1909.11090} \BibitemShut
  {NoStop}%
\bibitem [{\citenamefont {Di}\ and\ \citenamefont {Gong}(2018)}]{Gong:2017qlj}%
  \BibitemOpen
  \bibfield  {author} {\bibinfo {author} {\bibfnamefont {H.}~\bibnamefont
  {Di}}\ and\ \bibinfo {author} {\bibfnamefont {Y.}~\bibnamefont {Gong}},\
  }\bibinfo {title} {{Primordial black holes and second order gravitational
  waves from ultra-slow-roll inflation}},\ \href
  {https://doi.org/10.1088/1475-7516/2018/07/007} {J. Cosmol. Astropart. Phys.\
  \bibinfo {volume} {07}\bibfield  {year} {\bibinfo  {year} { (\textbf
  {2018})}\ }\bibfield  {pages} {\bibinfo  {pages} {007}},\ }\Eprint
  {https://arxiv.org/abs/1707.09578} {arXiv:1707.09578} \BibitemShut {NoStop}%
\bibitem [{\citenamefont {Motohashi}\ and\ \citenamefont
  {Hu}(2017)}]{Motohashi:2017kbs}%
  \BibitemOpen
  \bibfield  {author} {\bibinfo {author} {\bibfnamefont {H.}~\bibnamefont
  {Motohashi}}\ and\ \bibinfo {author} {\bibfnamefont {W.}~\bibnamefont {Hu}},\
  }\bibinfo {title} {{Primordial Black Holes and Slow-Roll Violation}},\ \href
  {https://doi.org/10.1103/PhysRevD.96.063503} {\bibfield  {journal} {\bibinfo
  {journal} {Phys. Rev. D}\ }\textbf {\bibinfo {volume} {96}},\ \bibinfo
  {pages} {063503} (\bibinfo {year} {2017})},\ \Eprint
  {https://arxiv.org/abs/1706.06784} {arXiv:1706.06784} \BibitemShut {NoStop}%
\bibitem [{\citenamefont {Lu}\ {\it et~al.}(2019)\citenamefont {Lu},
  \citenamefont {Gong}, \citenamefont {Yi},\ and\ \citenamefont
  {Zhang}}]{Lu:2019sti}%
  \BibitemOpen
  \bibfield  {author} {\bibinfo {author} {\bibfnamefont {Y.}~\bibnamefont
  {Lu}}, \bibinfo {author} {\bibfnamefont {Y.}~\bibnamefont {Gong}}, \bibinfo
  {author} {\bibfnamefont {Z.}~\bibnamefont {Yi}},\ and\ \bibinfo {author}
  {\bibfnamefont {F.}~\bibnamefont {Zhang}},\ }\bibinfo {title} {{Constraints
  on primordial curvature perturbations from primordial black hole dark matter
  and secondary gravitational waves}},\ \href
  {https://doi.org/10.1088/1475-7516/2019/12/031} {J. Cosmol. Astropart. Phys.\
  \bibinfo {volume} {12}\bibfield  {year} {\bibinfo  {year} { (\textbf
  {2019})}\ }\bibfield  {pages} {\bibinfo  {pages} {031}},\ }\Eprint
  {https://arxiv.org/abs/1907.11896} {arXiv:1907.11896} \BibitemShut {NoStop}%
\bibitem [{\citenamefont {Sato-Polito}\ {\it et~al.}(2019)\citenamefont
  {Sato-Polito}, \citenamefont {Kovetz},\ and\ \citenamefont
  {Kamionkowski}}]{Sato-Polito:2019hws}%
  \BibitemOpen
  \bibfield  {author} {\bibinfo {author} {\bibfnamefont {G.}~\bibnamefont
  {Sato-Polito}}, \bibinfo {author} {\bibfnamefont {E.~D.}\ \bibnamefont
  {Kovetz}},\ and\ \bibinfo {author} {\bibfnamefont {M.}~\bibnamefont
  {Kamionkowski}},\ }\bibinfo {title} {{Constraints on the primordial curvature
  power spectrum from primordial black holes}},\ \href
  {https://doi.org/10.1103/PhysRevD.100.063521} {\bibfield  {journal} {\bibinfo
   {journal} {Phys. Rev. D}\ }\textbf {\bibinfo {volume} {100}},\ \bibinfo
  {pages} {063521} (\bibinfo {year} {2019})},\ \Eprint
  {https://arxiv.org/abs/1904.10971} {arXiv:1904.10971} \BibitemShut {NoStop}%
\bibitem [{\citenamefont {Bugaev}\ and\ \citenamefont
  {Klimai}(2010)}]{Bugaev:2009zh}%
  \BibitemOpen
  \bibfield  {author} {\bibinfo {author} {\bibfnamefont {E.}~\bibnamefont
  {Bugaev}}\ and\ \bibinfo {author} {\bibfnamefont {P.}~\bibnamefont
  {Klimai}},\ }\bibinfo {title} {{Induced gravitational wave background and
  primordial black holes}},\ \href {https://doi.org/10.1103/PhysRevD.81.023517}
  {\bibfield  {journal} {\bibinfo  {journal} {Phys. Rev. D}\ }\textbf {\bibinfo
  {volume} {81}},\ \bibinfo {pages} {023517} (\bibinfo {year} {2010})},\
  \Eprint {https://arxiv.org/abs/0908.0664} {arXiv:0908.0664} \BibitemShut
  {NoStop}%
\bibitem [{\citenamefont {Bugaev}\ and\ \citenamefont
  {Klimai}(2011)}]{Bugaev:2010bb}%
  \BibitemOpen
  \bibfield  {author} {\bibinfo {author} {\bibfnamefont {E.}~\bibnamefont
  {Bugaev}}\ and\ \bibinfo {author} {\bibfnamefont {P.}~\bibnamefont
  {Klimai}},\ }\bibinfo {title} {{Constraints on the induced gravitational wave
  background from primordial black holes}},\ \href
  {https://doi.org/10.1103/PhysRevD.83.083521} {\bibfield  {journal} {\bibinfo
  {journal} {Phys. Rev. D}\ }\textbf {\bibinfo {volume} {83}},\ \bibinfo
  {pages} {083521} (\bibinfo {year} {2011})},\ \Eprint
  {https://arxiv.org/abs/1012.4697} {arXiv:1012.4697} \BibitemShut {NoStop}%
\bibitem [{\citenamefont {Alabidi}\ {\it et~al.}(2012)\citenamefont
  {Alabidi}, \citenamefont {Kohri}, \citenamefont {Sasaki},\ and\ \citenamefont
  {Sendouda}}]{Alabidi:2012ex}%
  \BibitemOpen
  \bibfield  {author} {\bibinfo {author} {\bibfnamefont {L.}~\bibnamefont
  {Alabidi}}, \bibinfo {author} {\bibfnamefont {K.}~\bibnamefont {Kohri}},
  \bibinfo {author} {\bibfnamefont {M.}~\bibnamefont {Sasaki}},\ and\ \bibinfo
  {author} {\bibfnamefont {Y.}~\bibnamefont {Sendouda}},\ }\bibinfo {title}
  {{Observable Spectra of Induced Gravitational Waves from Inflation}},\ \href
  {https://doi.org/10.1088/1475-7516/2012/09/017} {J. Cosmol. Astropart. Phys.\
  \bibinfo {volume} {1209}\bibfield  {year} {\bibinfo  {year} { (\textbf
  {2012})}\ }\bibfield  {pages} {\bibinfo  {pages} {017}},\ }\Eprint
  {https://arxiv.org/abs/1203.4663} {arXiv:1203.4663} \BibitemShut {NoStop}%
\bibitem [{\citenamefont {Inomata}\ \it
  {et~al.}(2017{\natexlab{b}})\citenamefont {Inomata}, \citenamefont
  {Kawasaki}, \citenamefont {Mukaida}, \citenamefont {Tada},\ and\
  \citenamefont {Yanagida}}]{Inomata:2016rbd}%
  \BibitemOpen
  \bibfield  {author} {\bibinfo {author} {\bibfnamefont {K.}~\bibnamefont
  {Inomata}}, \bibinfo {author} {\bibfnamefont {M.}~\bibnamefont {Kawasaki}},
  \bibinfo {author} {\bibfnamefont {K.}~\bibnamefont {Mukaida}}, \bibinfo
  {author} {\bibfnamefont {Y.}~\bibnamefont {Tada}},\ and\ \bibinfo {author}
  {\bibfnamefont {T.~T.}\ \bibnamefont {Yanagida}},\ }\bibinfo {title}
  {{Inflationary primordial black holes for the LIGO gravitational wave events
  and pulsar timing array experiments}},\ \href
  {https://doi.org/10.1103/PhysRevD.95.123510} {\bibfield  {journal} {\bibinfo
  {journal} {Phys. Rev. D}\ }\textbf {\bibinfo {volume} {95}},\ \bibinfo
  {pages} {123510} (\bibinfo {year} {2017}{\natexlab{b}})},\ \Eprint
  {https://arxiv.org/abs/1611.06130} {arXiv:1611.06130} \BibitemShut {NoStop}%
\bibitem [{\citenamefont {Drees}\ and\ \citenamefont
  {Xu}(2019)}]{Drees:2019xpp}%
  \BibitemOpen
  \bibfield  {author} {\bibinfo {author} {\bibfnamefont {M.}~\bibnamefont
  {Drees}}\ and\ \bibinfo {author} {\bibfnamefont {Y.}~\bibnamefont {Xu}},\
  }\bibinfo {title} {{Critical Higgs Inflation and Second Order Gravitational
  Wave Signatures}},\ \Eprint {https://arxiv.org/abs/1905.13581}
  {arXiv:1905.13581} \BibitemShut {NoStop}%
\bibitem [{\citenamefont {Inomata}\ \it
  {et~al.}(2019{\natexlab{a}})\citenamefont {Inomata}, \citenamefont {Kohri},
  \citenamefont {Nakama},\ and\ \citenamefont {Terada}}]{Inomata:2019ivs}%
  \BibitemOpen
  \bibfield  {author} {\bibinfo {author} {\bibfnamefont {K.}~\bibnamefont
  {Inomata}}, \bibinfo {author} {\bibfnamefont {K.}~\bibnamefont {Kohri}},
  \bibinfo {author} {\bibfnamefont {T.}~\bibnamefont {Nakama}},\ and\ \bibinfo
  {author} {\bibfnamefont {T.}~\bibnamefont {Terada}},\ }\bibinfo {title}
  {{Enhancement of Gravitational Waves Induced by Scalar Perturbations due to a
  Sudden Transition from an Early Matter Era to the Radiation Era}},\ \href
  {https://doi.org/10.1103/PhysRevD.100.043532} {\bibfield  {journal} {\bibinfo
   {journal} {Phys. Rev. D}\ }\textbf {\bibinfo {volume} {100}},\ \bibinfo
  {pages} {043532} (\bibinfo {year} {2019}{\natexlab{a}})},\ \Eprint
  {https://arxiv.org/abs/1904.12879} {arXiv:1904.12879} \BibitemShut {NoStop}%
\bibitem [{\citenamefont {Inomata}\ \it
  {et~al.}(2019{\natexlab{b}})\citenamefont {Inomata}, \citenamefont {Kohri},
  \citenamefont {Nakama},\ and\ \citenamefont {Terada}}]{Inomata:2019zqy}%
  \BibitemOpen
  \bibfield  {author} {\bibinfo {author} {\bibfnamefont {K.}~\bibnamefont
  {Inomata}}, \bibinfo {author} {\bibfnamefont {K.}~\bibnamefont {Kohri}},
  \bibinfo {author} {\bibfnamefont {T.}~\bibnamefont {Nakama}},\ and\ \bibinfo
  {author} {\bibfnamefont {T.}~\bibnamefont {Terada}},\ }\bibinfo {title}
  {{Gravitational Waves Induced by Scalar Perturbations during a Gradual
  Transition from an Early Matter Era to the Radiation Era}},\ \href
  {https://doi.org/10.1088/1475-7516/2019/10/071} {J. Cosmol. Astropart. Phys.\
  \bibinfo {volume} {10}\bibfield  {year} {\bibinfo  {year} { (\textbf
  {2019})}\ }\bibfield  {pages} {\bibinfo  {pages} {071}},\ }\Eprint
  {https://arxiv.org/abs/1904.12878} {arXiv:1904.12878} \BibitemShut {NoStop}%
\bibitem [{\citenamefont {De~Luca}\ {\it et~al.}(2019)\citenamefont
  {De~Luca}, \citenamefont {Desjacques}, \citenamefont {Franciolini},\ and\
  \citenamefont {Riotto}}]{DeLuca:2019llr}%
  \BibitemOpen
  \bibfield  {author} {\bibinfo {author} {\bibfnamefont {V.}~\bibnamefont
  {De~Luca}}, \bibinfo {author} {\bibfnamefont {V.}~\bibnamefont {Desjacques}},
  \bibinfo {author} {\bibfnamefont {G.}~\bibnamefont {Franciolini}},\ and\
  \bibinfo {author} {\bibfnamefont {A.}~\bibnamefont {Riotto}},\ }\bibinfo
  {title} {{Gravitational Waves from Peaks}},\ \href
  {https://doi.org/10.1088/1475-7516/2019/09/059} {J. Cosmol. Astropart. Phys.\
  \bibinfo {volume} {1909}\bibfield  {year} {\bibinfo  {year} { (\textbf
  {2019})}\ }\bibfield  {pages} {\bibinfo  {pages} {059}},\ }\Eprint
  {https://arxiv.org/abs/1905.13459} {arXiv:1905.13459} \BibitemShut {NoStop}%
\bibitem [{\citenamefont {Kuroyanagi}\ {\it et~al.}(2018)\citenamefont
  {Kuroyanagi}, \citenamefont {Chiba},\ and\ \citenamefont
  {Takahashi}}]{Kuroyanagi:2018csn}%
  \BibitemOpen
  \bibfield  {author} {\bibinfo {author} {\bibfnamefont {S.}~\bibnamefont
  {Kuroyanagi}}, \bibinfo {author} {\bibfnamefont {T.}~\bibnamefont {Chiba}},\
  and\ \bibinfo {author} {\bibfnamefont {T.}~\bibnamefont {Takahashi}},\
  }\bibinfo {title} {{Probing the Universe through the Stochastic Gravitational
  Wave Background}},\ \href {https://doi.org/10.1088/1475-7516/2018/11/038} {J.
  Cosmol. Astropart. Phys.\ \bibinfo {volume} {1811}\bibfield  {year} {\bibinfo
   {year} { (\textbf {2018})}\ }\bibfield  {pages} {\bibinfo  {pages} {038}},\
  }\Eprint {https://arxiv.org/abs/1807.00786} {arXiv:1807.00786} \BibitemShut
  {NoStop}%
\bibitem [{\citenamefont {Nakama}\ {\it et~al.}(2017)\citenamefont {Nakama},
  \citenamefont {Silk},\ and\ \citenamefont {Kamionkowski}}]{Nakama:2016gzw}%
  \BibitemOpen
  \bibfield  {author} {\bibinfo {author} {\bibfnamefont {T.}~\bibnamefont
  {Nakama}}, \bibinfo {author} {\bibfnamefont {J.}~\bibnamefont {Silk}},\ and\
  \bibinfo {author} {\bibfnamefont {M.}~\bibnamefont {Kamionkowski}},\
  }\bibinfo {title} {{Stochastic gravitational waves associated with the
  formation of primordial black holes}},\ \href
  {https://doi.org/10.1103/PhysRevD.95.043511} {\bibfield  {journal} {\bibinfo
  {journal} {Phys. Rev. D}\ }\textbf {\bibinfo {volume} {95}},\ \bibinfo
  {pages} {043511} (\bibinfo {year} {2017})},\ \Eprint
  {https://arxiv.org/abs/1612.06264} {arXiv:1612.06264} \BibitemShut {NoStop}%
\bibitem [{\citenamefont {Saito}\ and\ \citenamefont
  {Yokoyama}(2010)}]{Saito:2009jt}%
  \BibitemOpen
  \bibfield  {author} {\bibinfo {author} {\bibfnamefont {R.}~\bibnamefont
  {Saito}}\ and\ \bibinfo {author} {\bibfnamefont {J.}~\bibnamefont
  {Yokoyama}},\ }\bibinfo {title} {{Gravitational-Wave Constraints on the
  Abundance of Primordial Black Holes}},\ \href
  {https://doi.org/10.1143/PTP.126.351, 10.1143/PTP.123.867} {\bibfield
  {journal} {\bibinfo  {journal} {Prog. Theor. Phys.}\ }\textbf {\bibinfo
  {volume} {123}},\ \bibinfo {pages} {867} (\bibinfo {year} {2010})},\ \bibinfo
  {note} {[Erratum: Prog. Theor. Phys.126,351(2011)]},\ \Eprint
  {https://arxiv.org/abs/0912.5317} {arXiv:0912.5317} \BibitemShut {NoStop}%
\bibitem [{\citenamefont {Saito}\ and\ \citenamefont
  {Yokoyama}(2009)}]{Saito:2008jc}%
  \BibitemOpen
  \bibfield  {author} {\bibinfo {author} {\bibfnamefont {R.}~\bibnamefont
  {Saito}}\ and\ \bibinfo {author} {\bibfnamefont {J.}~\bibnamefont
  {Yokoyama}},\ }\bibinfo {title} {{Gravitational wave background as a probe of
  the primordial black hole abundance}},\ \href
  {https://doi.org/10.1103/PhysRevLett.102.161101,
  10.1103/PhysRevLett.107.069901} {\bibfield  {journal} {\bibinfo  {journal}
  {Phys. Rev. Lett.}\ }\textbf {\bibinfo {volume} {102}},\ \bibinfo {pages}
  {161101} (\bibinfo {year} {2009})},\ \bibinfo {note} {[Erratum: Phys. Rev.
  Lett.107,069901(2011)]},\ \Eprint {https://arxiv.org/abs/0812.4339}
  {arXiv:0812.4339} \BibitemShut {NoStop}%
\bibitem [{\citenamefont {Mollerach}\ {\it et~al.}(2004)\citenamefont
  {Mollerach}, \citenamefont {Harari},\ and\ \citenamefont
  {Matarrese}}]{Mollerach:2003nq}%
  \BibitemOpen
  \bibfield  {author} {\bibinfo {author} {\bibfnamefont {S.}~\bibnamefont
  {Mollerach}}, \bibinfo {author} {\bibfnamefont {D.}~\bibnamefont {Harari}},\
  and\ \bibinfo {author} {\bibfnamefont {S.}~\bibnamefont {Matarrese}},\
  }\bibinfo {title} {{CMB polarization from secondary vector and tensor
  modes}},\ \href {https://doi.org/10.1103/PhysRevD.69.063002} {\bibfield
  {journal} {\bibinfo  {journal} {Phys. Rev. D}\ }\textbf {\bibinfo {volume}
  {69}},\ \bibinfo {pages} {063002} (\bibinfo {year} {2004})},\ \Eprint
  {https://arxiv.org/abs/astro-ph/0310711} {arXiv:astro-ph/0310711}
  \BibitemShut {NoStop}%
\bibitem [{\citenamefont {Audley}\ {\it et~al.}(2017)\citenamefont {Audley}
  {\it et~al.}}]{Audley:2017drz}%
  \BibitemOpen
  \bibfield  {author} {\bibinfo {author} {\bibfnamefont {H.}~\bibnamefont
  {Audley}} {\it et~al.},\ }\bibinfo {title} {{Laser Interferometer Space
  Antenna}},\ \Eprint {https://arxiv.org/abs/1702.00786} {arXiv:1702.00786}
  \BibitemShut {NoStop}%
\bibitem [{\citenamefont {Danzmann}(1997)}]{Danzmann:1997hm}%
  \BibitemOpen
  \bibfield  {author} {\bibinfo {author} {\bibfnamefont {K.}~\bibnamefont
  {Danzmann}},\ }\bibinfo {title} {{LISA: An ESA cornerstone mission for a
  gravitational wave observatory}},\ \href
  {https://doi.org/10.1088/0264-9381/14/6/002} {\bibfield  {journal} {\bibinfo
  {journal} {Class. Quant. Grav.}\ }\textbf {\bibinfo {volume} {14}},\ \bibinfo
  {pages} {1399} (\bibinfo {year} {1997})}\BibitemShut {NoStop}%
\bibitem [{\citenamefont {Luo}\ {\it et~al.}(2016)\citenamefont {Luo} \it
  {et~al.}}]{Luo:2015ght}%
  \BibitemOpen
  \bibfield  {author} {\bibinfo {author} {\bibfnamefont {J.}~\bibnamefont
  {Luo}} {\it et~al.} (\bibinfo {collaboration} {TianQin}),\ }\bibinfo
  {title} {{TianQin: a space-borne gravitational wave detector}},\ \href
  {https://doi.org/10.1088/0264-9381/33/3/035010} {\bibfield  {journal}
  {\bibinfo  {journal} {Class. Quant. Grav.}\ }\textbf {\bibinfo {volume}
  {33}},\ \bibinfo {pages} {035010} (\bibinfo {year} {2016})},\ \Eprint
  {https://arxiv.org/abs/1512.02076} {arXiv:1512.02076} \BibitemShut {NoStop}%
\bibitem [{\citenamefont {Hu}\ and\ \citenamefont {Wu}(2017)}]{Hu:2017mde}%
  \BibitemOpen
  \bibfield  {author} {\bibinfo {author} {\bibfnamefont {W.-R.}\ \bibnamefont
  {Hu}}\ and\ \bibinfo {author} {\bibfnamefont {Y.-L.}\ \bibnamefont {Wu}},\
  }\bibinfo {title} {{The Taiji Program in Space for gravitational wave physics
  and the nature of gravity}},\ \href {https://doi.org/10.1093/nsr/nwx116}
  {\bibfield  {journal} {\bibinfo  {journal} {Natl. Sci. Rev.}\ }\textbf
  {\bibinfo {volume} {4}},\ \bibinfo {pages} {685} (\bibinfo {year}
  {2017})}\BibitemShut {NoStop}%
\bibitem [{\citenamefont {Hobbs}(2013)}]{Hobbs:2013aka}%
  \BibitemOpen
  \bibfield  {author} {\bibinfo {author} {\bibfnamefont {G.}~\bibnamefont
  {Hobbs}},\ }\bibinfo {title} {{The Parkes Pulsar Timing Array}},\ \href
  {https://doi.org/10.1088/0264-9381/30/22/224007} {\bibfield  {journal}
  {\bibinfo  {journal} {Class. Quant. Grav.}\ }\textbf {\bibinfo {volume}
  {30}},\ \bibinfo {pages} {224007} (\bibinfo {year} {2013})},\ \Eprint
  {https://arxiv.org/abs/1307.2629} {arXiv:1307.2629} \BibitemShut {NoStop}%
\bibitem [{\citenamefont {McLaughlin}(2013)}]{McLaughlin:2013ira}%
  \BibitemOpen
  \bibfield  {author} {\bibinfo {author} {\bibfnamefont {M.~A.}\ \bibnamefont
  {McLaughlin}},\ }\bibinfo {title} {{The North American Nanohertz Observatory
  for Gravitational Waves}},\ \href
  {https://doi.org/10.1088/0264-9381/30/22/224008} {\bibfield  {journal}
  {\bibinfo  {journal} {Class. Quant. Grav.}\ }\textbf {\bibinfo {volume}
  {30}},\ \bibinfo {pages} {224008} (\bibinfo {year} {2013})},\ \Eprint
  {https://arxiv.org/abs/1310.0758} {arXiv:1310.0758} \BibitemShut {NoStop}%
\bibitem [{\citenamefont {Hobbs}\ {\it et~al.}(2010)\citenamefont {Hobbs}
  {\it et~al.}}]{Hobbs:2009yy}%
  \BibitemOpen
  \bibfield  {author} {\bibinfo {author} {\bibfnamefont {G.}~\bibnamefont
  {Hobbs}} {\it et~al.},\ }\bibinfo {title}
  {{The international pulsar timing array project: using pulsars as a
  gravitational wave detector}},\ \href
  {https://doi.org/10.1088/0264-9381/27/8/084013} {\bibfield  {journal}
  {\bibinfo  {journal} {Class. Quant. Grav.}\ }\textbf {\bibinfo {volume}
  {27}},\ \bibinfo {pages} {084013} (\bibinfo {year} {2010})},\ \Eprint
  {https://arxiv.org/abs/0911.5206} {arXiv:0911.5206} \BibitemShut {NoStop}%
\bibitem [{\citenamefont {Kramer}\ and\ \citenamefont
  {Champion}(2013)}]{Kramer:2013kea}%
  \BibitemOpen
  \bibfield  {author} {\bibinfo {author} {\bibfnamefont {M.}~\bibnamefont
  {Kramer}}\ and\ \bibinfo {author} {\bibfnamefont {D.~J.}\ \bibnamefont
  {Champion}},\ }\bibinfo {title} {{The European Pulsar Timing Array and the
  Large European Array for Pulsars}},\ \href
  {https://doi.org/10.1088/0264-9381/30/22/224009} {\bibfield  {journal}
  {\bibinfo  {journal} {Class. Quant. Grav.}\ }\textbf {\bibinfo {volume}
  {30}},\ \bibinfo {pages} {224009} (\bibinfo {year} {2013})}\BibitemShut
  {NoStop}%
\bibitem [{\citenamefont {Moore}\ {\it et~al.}(2015)\citenamefont {Moore},
  \citenamefont {Cole},\ and\ \citenamefont {Berry}}]{Moore:2014lga}%
  \BibitemOpen
  \bibfield  {author} {\bibinfo {author} {\bibfnamefont {C.~J.}\ \bibnamefont
  {Moore}}, \bibinfo {author} {\bibfnamefont {R.~H.}\ \bibnamefont {Cole}},\
  and\ \bibinfo {author} {\bibfnamefont {C.~P.~L.}\ \bibnamefont {Berry}},\
  }\bibinfo {title} {{Gravitational-wave sensitivity curves}},\ \href
  {https://doi.org/10.1088/0264-9381/32/1/015014} {\bibfield  {journal}
  {\bibinfo  {journal} {Class. Quant. Grav.}\ }\textbf {\bibinfo {volume}
  {32}},\ \bibinfo {pages} {015014} (\bibinfo {year} {2015})},\ \Eprint
  {https://arxiv.org/abs/1408.0740} {arXiv:1408.0740} \BibitemShut {NoStop}%
\bibitem [{\citenamefont {Bezrukov}\ {\it et~al.}(2018)\citenamefont
  {Bezrukov}, \citenamefont {Pauly},\ and\ \citenamefont
  {Rubio}}]{Bezrukov:2017dyv}%
  \BibitemOpen
  \bibfield  {author} {\bibinfo {author} {\bibfnamefont {F.}~\bibnamefont
  {Bezrukov}}, \bibinfo {author} {\bibfnamefont {M.}~\bibnamefont {Pauly}},\
  and\ \bibinfo {author} {\bibfnamefont {J.}~\bibnamefont {Rubio}},\ }\bibinfo
  {title} {{On the robustness of the primordial power spectrum in renormalized
  Higgs inflation}},\ \href {https://doi.org/10.1088/1475-7516/2018/02/040} {J.
  Cosmol. Astropart. Phys.\ \bibinfo {volume} {02}\bibfield  {year} {\bibinfo
  {year} { (\textbf {2018})}\ }\bibfield  {pages} {\bibinfo  {pages} {040}},\
  }\Eprint {https://arxiv.org/abs/1706.05007} {arXiv:1706.05007} \BibitemShut
  {NoStop}%
\bibitem [{\citenamefont {Passaglia}\ {\it et~al.}(2019)\citenamefont
  {Passaglia}, \citenamefont {Hu},\ and\ \citenamefont
  {Motohashi}}]{Passaglia:2018ixg}%
  \BibitemOpen
  \bibfield  {author} {\bibinfo {author} {\bibfnamefont {S.}~\bibnamefont
  {Passaglia}}, \bibinfo {author} {\bibfnamefont {W.}~\bibnamefont {Hu}},\ and\
  \bibinfo {author} {\bibfnamefont {H.}~\bibnamefont {Motohashi}},\ }\bibinfo
  {title} {{Primordial black holes and local non-Gaussianity in canonical
  inflation}},\ \href {https://doi.org/10.1103/PhysRevD.99.043536} {\bibfield
  {journal} {\bibinfo  {journal} {Phys. Rev. D}\ }\textbf {\bibinfo {volume}
  {99}},\ \bibinfo {pages} {043536} (\bibinfo {year} {2019})},\ \Eprint
  {https://arxiv.org/abs/1812.08243} {arXiv:1812.08243} \BibitemShut {NoStop}%
\bibitem [{\citenamefont {Sasaki}\ {\it et~al.}(2018)\citenamefont {Sasaki},
  \citenamefont {Suyama}, \citenamefont {Tanaka},\ and\ \citenamefont
  {Yokoyama}}]{Sasaki:2018dmp}%
  \BibitemOpen
  \bibfield  {author} {\bibinfo {author} {\bibfnamefont {M.}~\bibnamefont
  {Sasaki}}, \bibinfo {author} {\bibfnamefont {T.}~\bibnamefont {Suyama}},
  \bibinfo {author} {\bibfnamefont {T.}~\bibnamefont {Tanaka}},\ and\ \bibinfo
  {author} {\bibfnamefont {S.}~\bibnamefont {Yokoyama}},\ }\bibinfo {title}
  {{Primordial black holes—perspectives in gravitational wave astronomy}},\
  \href {https://doi.org/10.1088/1361-6382/aaa7b4} {\bibfield  {journal}
  {\bibinfo  {journal} {Class. Quant. Grav.}\ }\textbf {\bibinfo {volume}
  {35}},\ \bibinfo {pages} {063001} (\bibinfo {year} {2018})},\ \Eprint
  {https://arxiv.org/abs/1801.05235} {arXiv:1801.05235} \BibitemShut {NoStop}%
\bibitem [{\citenamefont {Garcia-Bellido}\ and\ \citenamefont
  {Ruiz~Morales}(2017)}]{Garcia-Bellido:2017mdw}%
  \BibitemOpen
  \bibfield  {author} {\bibinfo {author} {\bibfnamefont {J.}~\bibnamefont
  {Garcia-Bellido}}\ and\ \bibinfo {author} {\bibfnamefont {E.}~\bibnamefont
  {Ruiz~Morales}},\ }\bibinfo {title} {{Primordial black holes from single
  field models of inflation}},\ \href
  {https://doi.org/10.1016/j.dark.2017.09.007} {\bibfield  {journal} {\bibinfo
  {journal} {Phys. Dark Univ.}\ }\textbf {\bibinfo {volume} {18}},\ \bibinfo
  {pages} {47} (\bibinfo {year} {2017})},\ \Eprint
  {https://arxiv.org/abs/1702.03901} {arXiv:1702.03901} \BibitemShut {NoStop}%
\bibitem [{\citenamefont {Ezquiaga}\ {\it et~al.}(2018)\citenamefont
  {Ezquiaga}, \citenamefont {Garcia-Bellido},\ and\ \citenamefont
  {Ruiz~Morales}}]{Ezquiaga:2017fvi}%
  \BibitemOpen
  \bibfield  {author} {\bibinfo {author} {\bibfnamefont {J.~M.}\ \bibnamefont
  {Ezquiaga}}, \bibinfo {author} {\bibfnamefont {J.}~\bibnamefont
  {Garcia-Bellido}},\ and\ \bibinfo {author} {\bibfnamefont {E.}~\bibnamefont
  {Ruiz~Morales}},\ }\bibinfo {title} {{Primordial Black Hole production in
  Critical Higgs Inflation}},\ \href
  {https://doi.org/10.1016/j.physletb.2017.11.039} {\bibfield  {journal}
  {\bibinfo  {journal} {Phys. Lett. B}\ }\textbf {\bibinfo {volume} {776}},\
  \bibinfo {pages} {345} (\bibinfo {year} {2018})},\ \Eprint
  {https://arxiv.org/abs/1705.04861} {arXiv:1705.04861} \BibitemShut {NoStop}%
\bibitem [{\citenamefont {Dimopoulos}(2017)}]{Dimopoulos:2017ged}%
  \BibitemOpen
  \bibfield  {author} {\bibinfo {author} {\bibfnamefont {K.}~\bibnamefont
  {Dimopoulos}},\ }\bibinfo {title} {{Ultra slow-roll inflation demystified}},\
  \href {https://doi.org/10.1016/j.physletb.2017.10.066} {\bibfield  {journal}
  {\bibinfo  {journal} {Phys. Lett. B}\ }\textbf {\bibinfo {volume} {775}},\
  \bibinfo {pages} {262} (\bibinfo {year} {2017})},\ \Eprint
  {https://arxiv.org/abs/1707.05644} {arXiv:1707.05644} \BibitemShut {NoStop}%
\bibitem [{\citenamefont {Espinosa}\ \it
  {et~al.}(2018{\natexlab{a}})\citenamefont {Espinosa}, \citenamefont {Racco},\
  and\ \citenamefont {Riotto}}]{Espinosa:2017sgp}%
  \BibitemOpen
  \bibfield  {author} {\bibinfo {author} {\bibfnamefont {J.~R.}\ \bibnamefont
  {Espinosa}}, \bibinfo {author} {\bibfnamefont {D.}~\bibnamefont {Racco}},\
  and\ \bibinfo {author} {\bibfnamefont {A.}~\bibnamefont {Riotto}},\ }\bibinfo
  {title} {{Cosmological Signature of the Standard Model Higgs Vacuum
  Instability: Primordial Black Holes as Dark Matter}},\ \href
  {https://doi.org/10.1103/PhysRevLett.120.121301} {\bibfield  {journal}
  {\bibinfo  {journal} {Phys. Rev. Lett.}\ }\textbf {\bibinfo {volume} {120}},\
  \bibinfo {pages} {121301} (\bibinfo {year} {2018}{\natexlab{a}})},\ \Eprint
  {https://arxiv.org/abs/1710.11196} {arXiv:1710.11196} \BibitemShut {NoStop}%
\bibitem [{\citenamefont {Espinosa}\ \it
  {et~al.}(2018{\natexlab{b}})\citenamefont {Espinosa}, \citenamefont {Racco},\
  and\ \citenamefont {Riotto}}]{Espinosa:2018eve}%
  \BibitemOpen
  \bibfield  {author} {\bibinfo {author} {\bibfnamefont {J.~R.}\ \bibnamefont
  {Espinosa}}, \bibinfo {author} {\bibfnamefont {D.}~\bibnamefont {Racco}},\
  and\ \bibinfo {author} {\bibfnamefont {A.}~\bibnamefont {Riotto}},\ }\bibinfo
  {title} {{A Cosmological Signature of the SM Higgs Instability: Gravitational
  Waves}},\ \href {https://doi.org/10.1088/1475-7516/2018/09/012} {J. Cosmol.
  Astropart. Phys.\ \bibinfo {volume} {09}\bibfield  {year} {\bibinfo  {year} {
  (\textbf {2018})}\ }\bibfield  {pages} {\bibinfo  {pages} {012}},\ }\Eprint
  {https://arxiv.org/abs/1804.07732} {arXiv:1804.07732} \BibitemShut {NoStop}%
\bibitem [{\citenamefont {Ballesteros}\ and\ \citenamefont
  {Taoso}(2018)}]{Ballesteros:2017fsr}%
  \BibitemOpen
  \bibfield  {author} {\bibinfo {author} {\bibfnamefont {G.}~\bibnamefont
  {Ballesteros}}\ and\ \bibinfo {author} {\bibfnamefont {M.}~\bibnamefont
  {Taoso}},\ }\bibinfo {title} {{Primordial black hole dark matter from single
  field inflation}},\ \href {https://doi.org/10.1103/PhysRevD.97.023501}
  {\bibfield  {journal} {\bibinfo  {journal} {Phys. Rev. D}\ }\textbf {\bibinfo
  {volume} {97}},\ \bibinfo {pages} {023501} (\bibinfo {year} {2018})},\
  \Eprint {https://arxiv.org/abs/1709.05565} {arXiv:1709.05565} \BibitemShut
  {NoStop}%
\bibitem [{\citenamefont {Germani}\ and\ \citenamefont
  {Prokopec}(2017)}]{Germani:2017bcs}%
  \BibitemOpen
  \bibfield  {author} {\bibinfo {author} {\bibfnamefont {C.}~\bibnamefont
  {Germani}}\ and\ \bibinfo {author} {\bibfnamefont {T.}~\bibnamefont
  {Prokopec}},\ }\bibinfo {title} {{On primordial black holes from an
  inflection point}},\ \href {https://doi.org/10.1016/j.dark.2017.09.001}
  {\bibfield  {journal} {\bibinfo  {journal} {Phys. Dark Univ.}\ }\textbf
  {\bibinfo {volume} {18}},\ \bibinfo {pages} {6} (\bibinfo {year} {2017})},\
  \Eprint {https://arxiv.org/abs/1706.04226} {arXiv:1706.04226} \BibitemShut
  {NoStop}%
\bibitem [{\citenamefont {Cheng}\ {\it et~al.}(2019)\citenamefont {Cheng},
  \citenamefont {Lee},\ and\ \citenamefont {Ng}}]{Cheng:2018qof}%
  \BibitemOpen
  \bibfield  {author} {\bibinfo {author} {\bibfnamefont {S.-L.}\ \bibnamefont
  {Cheng}}, \bibinfo {author} {\bibfnamefont {W.}~\bibnamefont {Lee}},\ and\
  \bibinfo {author} {\bibfnamefont {K.-W.}\ \bibnamefont {Ng}},\ }\bibinfo
  {title} {{Superhorizon curvature perturbation in ultraslow-roll inflation}},\
  \href {https://doi.org/10.1103/PhysRevD.99.063524} {\bibfield  {journal}
  {\bibinfo  {journal} {Phys. Rev. D}\ }\textbf {\bibinfo {volume} {99}},\
  \bibinfo {pages} {063524} (\bibinfo {year} {2019})},\ \Eprint
  {https://arxiv.org/abs/1811.10108} {arXiv:1811.10108} \BibitemShut {NoStop}%
\bibitem [{\citenamefont {Gao}\ and\ \citenamefont {Guo}(2018)}]{Gao:2018pvq}%
  \BibitemOpen
  \bibfield  {author} {\bibinfo {author} {\bibfnamefont {T.-J.}\ \bibnamefont
  {Gao}}\ and\ \bibinfo {author} {\bibfnamefont {Z.-K.}\ \bibnamefont {Guo}},\
  }\bibinfo {title} {{Primordial Black Hole Production in Inflationary Models
  of Supergravity with a Single Chiral Superfield}},\ \href
  {https://doi.org/10.1103/PhysRevD.98.063526} {\bibfield  {journal} {\bibinfo
  {journal} {Phys. Rev. D}\ }\textbf {\bibinfo {volume} {98}},\ \bibinfo
  {pages} {063526} (\bibinfo {year} {2018})},\ \Eprint
  {https://arxiv.org/abs/1806.09320} {arXiv:1806.09320} \BibitemShut {NoStop}%
\bibitem [{\citenamefont {Cicoli}\ {\it et~al.}(2018)\citenamefont {Cicoli},
  \citenamefont {Diaz},\ and\ \citenamefont {Pedro}}]{Cicoli:2018asa}%
  \BibitemOpen
  \bibfield  {author} {\bibinfo {author} {\bibfnamefont {M.}~\bibnamefont
  {Cicoli}}, \bibinfo {author} {\bibfnamefont {V.~A.}\ \bibnamefont {Diaz}},\
  and\ \bibinfo {author} {\bibfnamefont {F.~G.}\ \bibnamefont {Pedro}},\
  }\bibinfo {title} {{Primordial Black Holes from String Inflation}},\ \href
  {https://doi.org/10.1088/1475-7516/2018/06/034} {J. Cosmol. Astropart. Phys.\
  \bibinfo {volume} {06}\bibfield  {year} {\bibinfo  {year} { (\textbf
  {2018})}\ }\bibfield  {pages} {\bibinfo  {pages} {034}},\ }\Eprint
  {https://arxiv.org/abs/1803.02837} {arXiv:1803.02837} \BibitemShut {NoStop}%
\bibitem [{\citenamefont {Bhaumik}\ and\ \citenamefont
  {Jain}(2020)}]{Bhaumik:2019tvl}%
  \BibitemOpen
  \bibfield  {author} {\bibinfo {author} {\bibfnamefont {N.}~\bibnamefont
  {Bhaumik}}\ and\ \bibinfo {author} {\bibfnamefont {R.~K.}\ \bibnamefont
  {Jain}},\ }\bibinfo {title} {{Primordial black holes dark matter from
  inflection point models of inflation and the effects of reheating}},\ \href
  {https://doi.org/10.1088/1475-7516/2020/01/037} {J. Cosmol. Astropart. Phys.\
  \bibinfo {volume} {01}\bibfield  {year} {\bibinfo  {year} { (\textbf
  {2020})}\ }\bibfield  {pages} {\bibinfo  {pages} {037}},\ }\Eprint
  {https://arxiv.org/abs/1907.04125} {arXiv:1907.04125} \BibitemShut {NoStop}%
\bibitem [{\citenamefont {Xu}\ {\it et~al.}(2020)\citenamefont {Xu},
  \citenamefont {Liu}, \citenamefont {Gao},\ and\ \citenamefont
  {Guo}}]{Xu:2019bdp}%
  \BibitemOpen
  \bibfield  {author} {\bibinfo {author} {\bibfnamefont {W.-T.}\ \bibnamefont
  {Xu}}, \bibinfo {author} {\bibfnamefont {J.}~\bibnamefont {Liu}}, \bibinfo
  {author} {\bibfnamefont {T.-J.}\ \bibnamefont {Gao}},\ and\ \bibinfo {author}
  {\bibfnamefont {Z.-K.}\ \bibnamefont {Guo}},\ }\bibinfo {title}
  {{Gravitational waves from double-inflection-point inflation}},\ \href
  {https://doi.org/10.1103/PhysRevD.101.023505} {\bibfield  {journal} {\bibinfo
   {journal} {Phys. Rev. D}\ }\textbf {\bibinfo {volume} {101}},\ \bibinfo
  {pages} {023505} (\bibinfo {year} {2020})},\ \Eprint
  {https://arxiv.org/abs/1907.05213} {arXiv:1907.05213} \BibitemShut {NoStop}%
\bibitem [{\citenamefont {Fu}\ {\it et~al.}(2019)\citenamefont {Fu},
  \citenamefont {Wu},\ and\ \citenamefont {Yu}}]{Fu:2019ttf}%
  \BibitemOpen
  \bibfield  {author} {\bibinfo {author} {\bibfnamefont {C.}~\bibnamefont
  {Fu}}, \bibinfo {author} {\bibfnamefont {P.}~\bibnamefont {Wu}},\ and\
  \bibinfo {author} {\bibfnamefont {H.}~\bibnamefont {Yu}},\ }\bibinfo {title}
  {{Primordial Black Holes from Inflation with Nonminimal Derivative
  Coupling}},\ \href {https://doi.org/10.1103/PhysRevD.100.063532} {\bibfield
  {journal} {\bibinfo  {journal} {Phys. Rev. D}\ }\textbf {\bibinfo {volume}
  {100}},\ \bibinfo {pages} {063532} (\bibinfo {year} {2019})},\ \Eprint
  {https://arxiv.org/abs/1907.05042} {arXiv:1907.05042} \BibitemShut {NoStop}%
\bibitem [{\citenamefont {Passaglia}\ {\it et~al.}(2020)\citenamefont
  {Passaglia}, \citenamefont {Hu},\ and\ \citenamefont
  {Motohashi}}]{Passaglia:2019ueo}%
  \BibitemOpen
  \bibfield  {author} {\bibinfo {author} {\bibfnamefont {S.}~\bibnamefont
  {Passaglia}}, \bibinfo {author} {\bibfnamefont {W.}~\bibnamefont {Hu}},\ and\
  \bibinfo {author} {\bibfnamefont {H.}~\bibnamefont {Motohashi}},\ }\bibinfo
  {title} {{Primordial black holes as dark matter through Higgs field
  criticality}},\ \href {https://doi.org/10.1103/PhysRevD.101.123523}
  {\bibfield  {journal} {\bibinfo  {journal} {Phys. Rev. D}\ }\textbf {\bibinfo
  {volume} {101}},\ \bibinfo {pages} {123523} (\bibinfo {year} {2020})},\
  \Eprint {https://arxiv.org/abs/1912.02682} {arXiv:1912.02682} \BibitemShut
  {NoStop}%
\bibitem [{\citenamefont {Lin}\ {\it et~al.}(2020)\citenamefont {Lin},
  \citenamefont {Gao}, \citenamefont {Gong}, \citenamefont {Lu}, \citenamefont
  {Zhang},\ and\ \citenamefont {Zhang}}]{Lin:2020goi}%
  \BibitemOpen
  \bibfield  {author} {\bibinfo {author} {\bibfnamefont {J.}~\bibnamefont
  {Lin}}, \bibinfo {author} {\bibfnamefont {Q.}~\bibnamefont {Gao}}, \bibinfo
  {author} {\bibfnamefont {Y.}~\bibnamefont {Gong}}, \bibinfo {author}
  {\bibfnamefont {Y.}~\bibnamefont {Lu}}, \bibinfo {author} {\bibfnamefont
  {C.}~\bibnamefont {Zhang}},\ and\ \bibinfo {author} {\bibfnamefont
  {F.}~\bibnamefont {Zhang}},\ }\bibinfo {title} {{Primordial black holes and
  secondary gravitational waves from $k$ and $G$ inflation}},\ \href
  {https://doi.org/10.1103/PhysRevD.101.103515} {\bibfield  {journal} {\bibinfo
   {journal} {Phys. Rev. D}\ }\textbf {\bibinfo {volume} {101}},\ \bibinfo
  {pages} {103515} (\bibinfo {year} {2020})},\ \Eprint
  {https://arxiv.org/abs/2001.05909} {arXiv:2001.05909} \BibitemShut {NoStop}%
\bibitem [{\citenamefont {Palma}\ {\it et~al.}(2020)\citenamefont {Palma},
  \citenamefont {Sypsas},\ and\ \citenamefont {Zenteno}}]{Palma:2020ejf}%
  \BibitemOpen
  \bibfield  {author} {\bibinfo {author} {\bibfnamefont {G.~A.}\ \bibnamefont
  {Palma}}, \bibinfo {author} {\bibfnamefont {S.}~\bibnamefont {Sypsas}},\ and\
  \bibinfo {author} {\bibfnamefont {C.}~\bibnamefont {Zenteno}},\ }\bibinfo
  {title} {{Seeding primordial black holes in multi-field inflation}},\ \Eprint
  {https://arxiv.org/abs/2004.06106} {arXiv:2004.06106} \BibitemShut {NoStop}%
\bibitem [{\citenamefont {Braglia}\ {\it et~al.}(2020)\citenamefont
  {Braglia}, \citenamefont {Hazra}, \citenamefont {Finelli}, \citenamefont
  {Smoot}, \citenamefont {Sriramkumar},\ and\ \citenamefont
  {Starobinsky}}]{Braglia:2020eai}%
  \BibitemOpen
  \bibfield  {author} {\bibinfo {author} {\bibfnamefont {M.}~\bibnamefont
  {Braglia}}, \bibinfo {author} {\bibfnamefont {D.~K.}\ \bibnamefont {Hazra}},
  \bibinfo {author} {\bibfnamefont {F.}~\bibnamefont {Finelli}}, \bibinfo
  {author} {\bibfnamefont {G.~F.}\ \bibnamefont {Smoot}}, \bibinfo {author}
  {\bibfnamefont {L.}~\bibnamefont {Sriramkumar}},\ and\ \bibinfo {author}
  {\bibfnamefont {A.~A.}\ \bibnamefont {Starobinsky}},\ }\bibinfo {title}
  {{Generating PBHs and small-scale GWs in two-field models of inflation}},\
  \href@noop {} {\  (\bibinfo {year} {2020})},\ \Eprint
  {https://arxiv.org/abs/2005.02895} {arXiv:2005.02895} \BibitemShut {NoStop}%
\bibitem [{\citenamefont {Yi}\ {\it et~al.}(2020)\citenamefont {Yi},
  \citenamefont {Gong}, \citenamefont {Wang},\ and\ \citenamefont
  {Zhu}}]{Yi:2020kmq}%
  \BibitemOpen
  \bibfield  {author} {\bibinfo {author} {\bibfnamefont {Z.}~\bibnamefont
  {Yi}}, \bibinfo {author} {\bibfnamefont {Y.}~\bibnamefont {Gong}}, \bibinfo
  {author} {\bibfnamefont {B.}~\bibnamefont {Wang}},\ and\ \bibinfo {author}
  {\bibfnamefont {Z.-h.}\ \bibnamefont {Zhu}},\ }\bibinfo {title} {{Primordial
  Black Holes and Secondary Gravitational Waves from Higgs field}},\ \href@noop
  {} {\  (\bibinfo {year} {2020})},\ \Eprint {https://arxiv.org/abs/2007.09957}
  {arXiv:2007.09957} \BibitemShut {NoStop}%
\bibitem [{\citenamefont {Ananda}\ {\it et~al.}(2007)\citenamefont {Ananda},
  \citenamefont {Clarkson},\ and\ \citenamefont {Wands}}]{Ananda:2006af}%
  \BibitemOpen
  \bibfield  {author} {\bibinfo {author} {\bibfnamefont {K.~N.}\ \bibnamefont
  {Ananda}}, \bibinfo {author} {\bibfnamefont {C.}~\bibnamefont {Clarkson}},\
  and\ \bibinfo {author} {\bibfnamefont {D.}~\bibnamefont {Wands}},\ }\bibinfo
  {title} {{The Cosmological gravitational wave background from primordial
  density perturbations}},\ \href {https://doi.org/10.1103/PhysRevD.75.123518}
  {\bibfield  {journal} {\bibinfo  {journal} {Phys. Rev. D}\ }\textbf {\bibinfo
  {volume} {75}},\ \bibinfo {pages} {123518} (\bibinfo {year} {2007})},\
  \Eprint {https://arxiv.org/abs/gr-qc/0612013} {arXiv:gr-qc/0612013}
  \BibitemShut {NoStop}%
\bibitem [{\citenamefont {Namba}\ {\it et~al.}(2016)\citenamefont {Namba},
  \citenamefont {Peloso}, \citenamefont {Shiraishi}, \citenamefont {Sorbo},\
  and\ \citenamefont {Unal}}]{Namba:2015gja}%
  \BibitemOpen
  \bibfield  {author} {\bibinfo {author} {\bibfnamefont {R.}~\bibnamefont
  {Namba}}, \bibinfo {author} {\bibfnamefont {M.}~\bibnamefont {Peloso}},
  \bibinfo {author} {\bibfnamefont {M.}~\bibnamefont {Shiraishi}}, \bibinfo
  {author} {\bibfnamefont {L.}~\bibnamefont {Sorbo}},\ and\ \bibinfo {author}
  {\bibfnamefont {C.}~\bibnamefont {Unal}},\ }\bibinfo {title}
  {{Scale-dependent gravitational waves from a rolling axion}},\ \href
  {https://doi.org/10.1088/1475-7516/2016/01/041} {J. Cosmol. Astropart. Phys.\
  \bibinfo {volume} {1601}\bibfield  {year} {\bibinfo  {year} { (\textbf
  {2016})}\ }\bibfield  {pages} {\bibinfo  {pages} {041}},\ }\Eprint
  {https://arxiv.org/abs/1509.07521} {arXiv:1509.07521} \BibitemShut {NoStop}%
\bibitem [{\citenamefont {Nakama}\ and\ \citenamefont
  {Suyama}(2015)}]{Nakama:2015nea}%
  \BibitemOpen
  \bibfield  {author} {\bibinfo {author} {\bibfnamefont {T.}~\bibnamefont
  {Nakama}}\ and\ \bibinfo {author} {\bibfnamefont {T.}~\bibnamefont
  {Suyama}},\ }\bibinfo {title} {{Primordial black holes as a novel probe of
  primordial gravitational waves}},\ \href
  {https://doi.org/10.1103/PhysRevD.92.121304} {\bibfield  {journal} {\bibinfo
  {journal} {Phys. Rev. D}\ }\textbf {\bibinfo {volume} {92}},\ \bibinfo
  {pages} {121304} (\bibinfo {year} {2015})},\ \Eprint
  {https://arxiv.org/abs/1506.05228} {arXiv:1506.05228} \BibitemShut {NoStop}%
\bibitem [{\citenamefont {Nakama}\ and\ \citenamefont
  {Suyama}(2016)}]{Nakama:2016enz}%
  \BibitemOpen
  \bibfield  {author} {\bibinfo {author} {\bibfnamefont {T.}~\bibnamefont
  {Nakama}}\ and\ \bibinfo {author} {\bibfnamefont {T.}~\bibnamefont
  {Suyama}},\ }\bibinfo {title} {{Primordial black holes as a novel probe of
  primordial gravitational waves. II: Detailed analysis}},\ \href
  {https://doi.org/10.1103/PhysRevD.94.043507} {\bibfield  {journal} {\bibinfo
  {journal} {Phys. Rev. D}\ }\textbf {\bibinfo {volume} {94}},\ \bibinfo
  {pages} {043507} (\bibinfo {year} {2016})},\ \Eprint
  {https://arxiv.org/abs/1605.04482} {arXiv:1605.04482} \BibitemShut {NoStop}%
\bibitem [{\citenamefont {Bartolo}\ \it
  {et~al.}(2019{\natexlab{a}})\citenamefont {Bartolo}, \citenamefont {De~Luca},
  \citenamefont {Franciolini}, \citenamefont {Lewis}, \citenamefont {Peloso},\
  and\ \citenamefont {Riotto}}]{Bartolo:2018evs}%
  \BibitemOpen
  \bibfield  {author} {\bibinfo {author} {\bibfnamefont {N.}~\bibnamefont
  {Bartolo}}, \bibinfo {author} {\bibfnamefont {V.}~\bibnamefont {De~Luca}},
  \bibinfo {author} {\bibfnamefont {G.}~\bibnamefont {Franciolini}}, \bibinfo
  {author} {\bibfnamefont {A.}~\bibnamefont {Lewis}}, \bibinfo {author}
  {\bibfnamefont {M.}~\bibnamefont {Peloso}},\ and\ \bibinfo {author}
  {\bibfnamefont {A.}~\bibnamefont {Riotto}},\ }\bibinfo {title} {{Primordial
  Black Hole Dark Matter: LISA Serendipity}},\ \href
  {https://doi.org/10.1103/PhysRevLett.122.211301} {\bibfield  {journal}
  {\bibinfo  {journal} {Phys. Rev. Lett.}\ }\textbf {\bibinfo {volume} {122}},\
  \bibinfo {pages} {211301} (\bibinfo {year} {2019}{\natexlab{a}})},\ \Eprint
  {https://arxiv.org/abs/1810.12218} {arXiv:1810.12218} \BibitemShut {NoStop}%
\bibitem [{\citenamefont {Byrnes}\ {\it et~al.}(2019)\citenamefont {Byrnes},
  \citenamefont {Cole},\ and\ \citenamefont {Patil}}]{Byrnes:2018txb}%
  \BibitemOpen
  \bibfield  {author} {\bibinfo {author} {\bibfnamefont {C.~T.}\ \bibnamefont
  {Byrnes}}, \bibinfo {author} {\bibfnamefont {P.~S.}\ \bibnamefont {Cole}},\
  and\ \bibinfo {author} {\bibfnamefont {S.~P.}\ \bibnamefont {Patil}},\
  }\bibinfo {title} {{Steepest growth of the power spectrum and primordial
  black holes}},\ \href {https://doi.org/10.1088/1475-7516/2019/06/028} {J.
  Cosmol. Astropart. Phys.\ \bibinfo {volume} {1906}\bibfield  {year} {\bibinfo
   {year} { (\textbf {2019})}\ }\bibfield  {pages} {\bibinfo  {pages} {028}},\
  }\Eprint {https://arxiv.org/abs/1811.11158} {arXiv:1811.11158} \BibitemShut
  {NoStop}%
\bibitem [{\citenamefont {Orlofsky}\ {\it et~al.}(2017)\citenamefont
  {Orlofsky}, \citenamefont {Pierce},\ and\ \citenamefont
  {Wells}}]{Orlofsky:2016vbd}%
  \BibitemOpen
  \bibfield  {author} {\bibinfo {author} {\bibfnamefont {N.}~\bibnamefont
  {Orlofsky}}, \bibinfo {author} {\bibfnamefont {A.}~\bibnamefont {Pierce}},\
  and\ \bibinfo {author} {\bibfnamefont {J.~D.}\ \bibnamefont {Wells}},\
  }\bibinfo {title} {{Inflationary theory and pulsar timing investigations of
  primordial black holes and gravitational waves}},\ \href
  {https://doi.org/10.1103/PhysRevD.95.063518} {\bibfield  {journal} {\bibinfo
  {journal} {Phys. Rev. D}\ }\textbf {\bibinfo {volume} {95}},\ \bibinfo
  {pages} {063518} (\bibinfo {year} {2017})},\ \Eprint
  {https://arxiv.org/abs/1612.05279} {arXiv:1612.05279} \BibitemShut {NoStop}%
\bibitem [{\citenamefont {Garcia-Bellido}\ {\it et~al.}(2017)\citenamefont
  {Garcia-Bellido}, \citenamefont {Peloso},\ and\ \citenamefont
  {Unal}}]{Garcia-Bellido:2017aan}%
  \BibitemOpen
  \bibfield  {author} {\bibinfo {author} {\bibfnamefont {J.}~\bibnamefont
  {Garcia-Bellido}}, \bibinfo {author} {\bibfnamefont {M.}~\bibnamefont
  {Peloso}},\ and\ \bibinfo {author} {\bibfnamefont {C.}~\bibnamefont {Unal}},\
  }\bibinfo {title} {{Gravitational Wave signatures of inflationary models from
  Primordial Black Hole Dark Matter}},\ \href
  {https://doi.org/10.1088/1475-7516/2017/09/013} {J. Cosmol. Astropart. Phys.\
  \bibinfo {volume} {09}\bibfield  {year} {\bibinfo  {year} { (\textbf
  {2017})}\ }\bibfield  {pages} {\bibinfo  {pages} {013}},\ }\Eprint
  {https://arxiv.org/abs/1707.02441} {arXiv:1707.02441} \BibitemShut {NoStop}%
\bibitem [{\citenamefont {Kohri}\ and\ \citenamefont
  {Terada}(2018)}]{Kohri:2018awv}%
  \BibitemOpen
  \bibfield  {author} {\bibinfo {author} {\bibfnamefont {K.}~\bibnamefont
  {Kohri}}\ and\ \bibinfo {author} {\bibfnamefont {T.}~\bibnamefont {Terada}},\
  }\bibinfo {title} {{Semianalytic calculation of gravitational wave spectrum
  nonlinearly induced from primordial curvature perturbations}},\ \href
  {https://doi.org/10.1103/PhysRevD.97.123532} {\bibfield  {journal} {\bibinfo
  {journal} {Phys. Rev. D}\ }\textbf {\bibinfo {volume} {97}},\ \bibinfo
  {pages} {123532} (\bibinfo {year} {2018})},\ \Eprint
  {https://arxiv.org/abs/1804.08577} {arXiv:1804.08577} \BibitemShut {NoStop}%
\bibitem [{\citenamefont {Bartolo}\ \it
  {et~al.}(2019{\natexlab{b}})\citenamefont {Bartolo}, \citenamefont {De~Luca},
  \citenamefont {Franciolini}, \citenamefont {Peloso}, \citenamefont {Racco},\
  and\ \citenamefont {Riotto}}]{Bartolo:2018rku}%
  \BibitemOpen
  \bibfield  {author} {\bibinfo {author} {\bibfnamefont {N.}~\bibnamefont
  {Bartolo}}, \bibinfo {author} {\bibfnamefont {V.}~\bibnamefont {De~Luca}},
  \bibinfo {author} {\bibfnamefont {G.}~\bibnamefont {Franciolini}}, \bibinfo
  {author} {\bibfnamefont {M.}~\bibnamefont {Peloso}}, \bibinfo {author}
  {\bibfnamefont {D.}~\bibnamefont {Racco}},\ and\ \bibinfo {author}
  {\bibfnamefont {A.}~\bibnamefont {Riotto}},\ }\bibinfo {title} {{Testing
  primordial black holes as dark matter with LISA}},\ \href
  {https://doi.org/10.1103/PhysRevD.99.103521} {\bibfield  {journal} {\bibinfo
  {journal} {Phys. Rev. D}\ }\textbf {\bibinfo {volume} {99}},\ \bibinfo
  {pages} {103521} (\bibinfo {year} {2019}{\natexlab{b}})},\ \Eprint
  {https://arxiv.org/abs/1810.12224} {arXiv:1810.12224} \BibitemShut {NoStop}%
\bibitem [{\citenamefont {Cai}\ {\it et~al.}(2019)\citenamefont {Cai},
  \citenamefont {Pi},\ and\ \citenamefont {Sasaki}}]{Cai:2018dig}%
  \BibitemOpen
  \bibfield  {author} {\bibinfo {author} {\bibfnamefont {R.-G.}\ \bibnamefont
  {Cai}}, \bibinfo {author} {\bibfnamefont {S.}~\bibnamefont {Pi}},\ and\
  \bibinfo {author} {\bibfnamefont {M.}~\bibnamefont {Sasaki}},\ }\bibinfo
  {title} {{Gravitational Waves Induced by non-Gaussian Scalar
  Perturbations}},\ \href {https://doi.org/10.1103/PhysRevLett.122.201101}
  {\bibfield  {journal} {\bibinfo  {journal} {Phys. Rev. Lett.}\ }\textbf
  {\bibinfo {volume} {122}},\ \bibinfo {pages} {201101} (\bibinfo {year}
  {2019})},\ \Eprint {https://arxiv.org/abs/1810.11000} {arXiv:1810.11000}
  \BibitemShut {NoStop}%
\bibitem [{\citenamefont {Unal}(2019)}]{Unal:2018yaa}%
  \BibitemOpen
  \bibfield  {author} {\bibinfo {author} {\bibfnamefont {C.}~\bibnamefont
  {Unal}},\ }\bibinfo {title} {{Imprints of Primordial Non-Gaussianity on
  Gravitational Wave Spectrum}},\ \href
  {https://doi.org/10.1103/PhysRevD.99.041301} {\bibfield  {journal} {\bibinfo
  {journal} {Phys. Rev. D}\ }\textbf {\bibinfo {volume} {99}},\ \bibinfo
  {pages} {041301} (\bibinfo {year} {2019})},\ \Eprint
  {https://arxiv.org/abs/1811.09151} {arXiv:1811.09151} \BibitemShut {NoStop}%
\bibitem [{\citenamefont {Inomata}\ and\ \citenamefont
  {Nakama}(2019)}]{Inomata:2018epa}%
  \BibitemOpen
  \bibfield  {author} {\bibinfo {author} {\bibfnamefont {K.}~\bibnamefont
  {Inomata}}\ and\ \bibinfo {author} {\bibfnamefont {T.}~\bibnamefont
  {Nakama}},\ }\bibinfo {title} {{Gravitational waves induced by scalar
  perturbations as probes of the small-scale primordial spectrum}},\ \href
  {https://doi.org/10.1103/PhysRevD.99.043511} {\bibfield  {journal} {\bibinfo
  {journal} {Phys. Rev. D}\ }\textbf {\bibinfo {volume} {99}},\ \bibinfo
  {pages} {043511} (\bibinfo {year} {2019})},\ \Eprint
  {https://arxiv.org/abs/1812.00674} {arXiv:1812.00674} \BibitemShut {NoStop}%
\bibitem [{\citenamefont {Germani}\ and\ \citenamefont
  {Musco}(2019)}]{Germani:2018jgr}%
  \BibitemOpen
  \bibfield  {author} {\bibinfo {author} {\bibfnamefont {C.}~\bibnamefont
  {Germani}}\ and\ \bibinfo {author} {\bibfnamefont {I.}~\bibnamefont
  {Musco}},\ }\bibinfo {title} {{Abundance of Primordial Black Holes Depends on
  the Shape of the Inflationary Power Spectrum}},\ \href
  {https://doi.org/10.1103/PhysRevLett.122.141302} {\bibfield  {journal}
  {\bibinfo  {journal} {Phys. Rev. Lett.}\ }\textbf {\bibinfo {volume} {122}},\
  \bibinfo {pages} {141302} (\bibinfo {year} {2019})},\ \Eprint
  {https://arxiv.org/abs/1805.04087} {arXiv:1805.04087} \BibitemShut {NoStop}%
\bibitem [{\citenamefont {Pi}\ and\ \citenamefont {Sasaki}(2020)}]{Pi:2020otn}%
  \BibitemOpen
  \bibfield  {author} {\bibinfo {author} {\bibfnamefont {S.}~\bibnamefont
  {Pi}}\ and\ \bibinfo {author} {\bibfnamefont {M.}~\bibnamefont {Sasaki}},\
  }\bibinfo {title} {{Gravitational Waves Induced by Scalar Perturbations with
  a Lognormal Peak}},\ \Eprint {https://arxiv.org/abs/2005.12306}
  {arXiv:2005.12306} \BibitemShut {NoStop}%
\bibitem [{\citenamefont {Baumann}\ {\it et~al.}(2007)\citenamefont
  {Baumann}, \citenamefont {Steinhardt}, \citenamefont {Takahashi},\ and\
  \citenamefont {Ichiki}}]{Baumann:2007zm}%
  \BibitemOpen
  \bibfield  {author} {\bibinfo {author} {\bibfnamefont {D.}~\bibnamefont
  {Baumann}}, \bibinfo {author} {\bibfnamefont {P.~J.}\ \bibnamefont
  {Steinhardt}}, \bibinfo {author} {\bibfnamefont {K.}~\bibnamefont
  {Takahashi}},\ and\ \bibinfo {author} {\bibfnamefont {K.}~\bibnamefont
  {Ichiki}},\ }\bibinfo {title} {{Gravitational Wave Spectrum Induced by
  Primordial Scalar Perturbations}},\ \href
  {https://doi.org/10.1103/PhysRevD.76.084019} {\bibfield  {journal} {\bibinfo
  {journal} {Phys. Rev. D}\ }\textbf {\bibinfo {volume} {76}},\ \bibinfo
  {pages} {084019} (\bibinfo {year} {2007})},\ \Eprint
  {https://arxiv.org/abs/hep-th/0703290} {arXiv:hep-th/0703290} \BibitemShut
  {NoStop}%
\bibitem [{\citenamefont {Hiramatsu}\ {\it et~al.}(2014)\citenamefont
  {Hiramatsu}, \citenamefont {Kawasaki},\ and\ \citenamefont
  {Saikawa}}]{Hiramatsu:2013qaa}%
  \BibitemOpen
  \bibfield  {author} {\bibinfo {author} {\bibfnamefont {T.}~\bibnamefont
  {Hiramatsu}}, \bibinfo {author} {\bibfnamefont {M.}~\bibnamefont
  {Kawasaki}},\ and\ \bibinfo {author} {\bibfnamefont {K.}~\bibnamefont
  {Saikawa}},\ }\bibinfo {title} {{On the estimation of gravitational wave
  spectrum from cosmic domain walls}},\ \href
  {https://doi.org/10.1088/1475-7516/2014/02/031} {J. Cosmol. Astropart. Phys.\
  \bibinfo {volume} {02}\bibfield  {year} {\bibinfo  {year} { (\textbf
  {2014})}\ }\bibfield  {pages} {\bibinfo  {pages} {031}},\ }\Eprint
  {https://arxiv.org/abs/1309.5001} {arXiv:1309.5001} \BibitemShut {NoStop}%
\bibitem [{\citenamefont {Bullock}\ and\ \citenamefont
  {Primack}(1997)}]{Bullock:1996at}%
  \BibitemOpen
  \bibfield  {author} {\bibinfo {author} {\bibfnamefont {J.~S.}\ \bibnamefont
  {Bullock}}\ and\ \bibinfo {author} {\bibfnamefont {J.~R.}\ \bibnamefont
  {Primack}},\ }\bibinfo {title} {{NonGaussian fluctuations and primordial
  black holes from inflation}},\ \href
  {https://doi.org/10.1103/PhysRevD.55.7423} {\bibfield  {journal} {\bibinfo
  {journal} {Phys. Rev. D}\ }\textbf {\bibinfo {volume} {55}},\ \bibinfo
  {pages} {7423} (\bibinfo {year} {1997})},\ \Eprint
  {https://arxiv.org/abs/astro-ph/9611106} {arXiv:astro-ph/9611106}
  \BibitemShut {NoStop}%
\bibitem [{\citenamefont {Saito}\ {\it et~al.}(2008)\citenamefont {Saito},
  \citenamefont {Yokoyama},\ and\ \citenamefont {Nagata}}]{Saito:2008em}%
  \BibitemOpen
  \bibfield  {author} {\bibinfo {author} {\bibfnamefont {R.}~\bibnamefont
  {Saito}}, \bibinfo {author} {\bibfnamefont {J.}~\bibnamefont {Yokoyama}},\
  and\ \bibinfo {author} {\bibfnamefont {R.}~\bibnamefont {Nagata}},\ }\bibinfo
  {title} {{Single-field inflation, anomalous enhancement of superhorizon
  fluctuations, and non-Gaussianity in primordial black hole formation}},\
  \href {https://doi.org/10.1088/1475-7516/2008/06/024} {J. Cosmol. Astropart.
  Phys.\ \bibinfo {volume} {0806}\bibfield  {year} {\bibinfo  {year} { (\textbf
  {2008})}\ }\bibfield  {pages} {\bibinfo  {pages} {024}},\ }\Eprint
  {https://arxiv.org/abs/0804.3470} {arXiv:0804.3470} \BibitemShut {NoStop}%
\bibitem [{\citenamefont {Kadota}\ {\it et~al.}(2015)\citenamefont {Kadota},
  \citenamefont {Kawasaki},\ and\ \citenamefont {Saikawa}}]{Kadota:2015dza}%
  \BibitemOpen
  \bibfield  {author} {\bibinfo {author} {\bibfnamefont {K.}~\bibnamefont
  {Kadota}}, \bibinfo {author} {\bibfnamefont {M.}~\bibnamefont {Kawasaki}},\
  and\ \bibinfo {author} {\bibfnamefont {K.}~\bibnamefont {Saikawa}},\
  }\bibinfo {title} {{Gravitational waves from domain walls in the
  next-to-minimal supersymmetric standard model}},\ \href
  {https://doi.org/10.1088/1475-7516/2015/10/041} {J. Cosmol. Astropart. Phys.\
  \bibinfo {volume} {10}\bibfield  {year} {\bibinfo  {year} { (\textbf
  {2015})}\ }\bibfield  {pages} {\bibinfo  {pages} {041}},\ }\Eprint
  {https://arxiv.org/abs/1503.06998} {arXiv:1503.06998} \BibitemShut {NoStop}%
\bibitem [{\citenamefont {Dalianis}\ {\it et~al.}(2019)\citenamefont
  {Dalianis}, \citenamefont {Kehagias},\ and\ \citenamefont
  {Tringas}}]{Dalianis:2018frf}%
  \BibitemOpen
  \bibfield  {author} {\bibinfo {author} {\bibfnamefont {I.}~\bibnamefont
  {Dalianis}}, \bibinfo {author} {\bibfnamefont {A.}~\bibnamefont {Kehagias}},\
  and\ \bibinfo {author} {\bibfnamefont {G.}~\bibnamefont {Tringas}},\
  }\bibinfo {title} {{Primordial black holes from $\alpha$-attractors}},\ \href
  {https://doi.org/10.1088/1475-7516/2019/01/037} {J. Cosmol. Astropart. Phys.\
  \bibinfo {volume} {1901}\bibfield  {year} {\bibinfo  {year} { (\textbf
  {2019})}\ }\bibfield  {pages} {\bibinfo  {pages} {037}},\ }\Eprint
  {https://arxiv.org/abs/1805.09483} {arXiv:1805.09483} \BibitemShut {NoStop}%
\bibitem [{\citenamefont {Tada}\ and\ \citenamefont
  {Yokoyama}(2019)}]{Tada:2019amh}%
  \BibitemOpen
  \bibfield  {author} {\bibinfo {author} {\bibfnamefont {Y.}~\bibnamefont
  {Tada}}\ and\ \bibinfo {author} {\bibfnamefont {S.}~\bibnamefont
  {Yokoyama}},\ }\bibinfo {title} {{Primordial black hole tower: Dark matter,
  earth-mass, and LIGO black holes}},\ \href
  {https://doi.org/10.1103/PhysRevD.100.023537} {\bibfield  {journal} {\bibinfo
   {journal} {Phys. Rev. D}\ }\textbf {\bibinfo {volume} {100}},\ \bibinfo
  {pages} {023537} (\bibinfo {year} {2019})},\ \Eprint
  {https://arxiv.org/abs/1904.10298} {arXiv:1904.10298} \BibitemShut {NoStop}%
\bibitem [{\citenamefont {Fu}\ {\it et~al.}(2020)\citenamefont {Fu},
  \citenamefont {Wu},\ and\ \citenamefont {Yu}}]{Fu:2019vqc}%
  \BibitemOpen
  \bibfield  {author} {\bibinfo {author} {\bibfnamefont {C.}~\bibnamefont
  {Fu}}, \bibinfo {author} {\bibfnamefont {P.}~\bibnamefont {Wu}},\ and\
  \bibinfo {author} {\bibfnamefont {H.}~\bibnamefont {Yu}},\ }\bibinfo {title}
  {{Scalar induced gravitational waves in inflation with gravitationally
  enhanced friction}},\ \href {https://doi.org/10.1103/PhysRevD.101.023529}
  {\bibfield  {journal} {\bibinfo  {journal} {Phys. Rev. D}\ }\textbf {\bibinfo
  {volume} {101}},\ \bibinfo {pages} {023529} (\bibinfo {year} {2020})},\
  \Eprint {https://arxiv.org/abs/1912.05927} {arXiv:1912.05927} \BibitemShut
  {NoStop}%
\bibitem [{\citenamefont {Akrami}\ {\it et~al.}(2018)\citenamefont {Akrami}
  {\it et~al.}}]{Akrami:2018odb}%
  \BibitemOpen
  \bibfield  {author} {\bibinfo {author} {\bibfnamefont {Y.}~\bibnamefont
  {Akrami}} {\it et~al.} (\bibinfo {collaboration} {Planck}),\ }\bibinfo
  {title} {{Planck 2018 results. X. Constraints on inflation}},\ \Eprint
  {https://arxiv.org/abs/1807.06211} {arXiv:1807.06211} \BibitemShut {NoStop}%
\bibitem [{\citenamefont {Aghanim}\ {\it et~al.}(2018)\citenamefont {Aghanim}
  {\it et~al.}}]{Aghanim:2018eyx}%
  \BibitemOpen
  \bibfield  {author} {\bibinfo {author} {\bibfnamefont {N.}~\bibnamefont
  {Aghanim}} {\it et~al.} (\bibinfo {collaboration} {Planck}),\ }\bibinfo
  {title} {{Planck 2018 results. VI. Cosmological parameters}},\ \Eprint
  {https://arxiv.org/abs/1807.06209} {arXiv:1807.06209} \BibitemShut {NoStop}%
\end{thebibliography}

%

\end{document}